\documentclass{elsart}
\def\newline{\relax\ifhmode\null\hfil\break\else\nonhmodeerr@\newline\fi}
\def\frac#1#2{{#1\over#2}}
\def\text#1{{\hbox{\rm #1}}}
\def\flushpar{{\par \noindent}}

\newcommand{\beq}{\begin{equation}}
\newcommand{\eeq}{\end{equation}}
\newcommand{\bea}{\begin{eqnarray}}
\newcommand{\eea}{\end{eqnarray}}
\def\Id{ \mbox{1\hspace{-1.2mm}I} }
\def\BE{\begin{equation}}
\def\EE{\end{equation}}
\def\BA{\begin{eqnarray}}
\def\EA{\end{eqnarray}}
\def\BAN{\begin{eqnarray*}}
\def\EAN{\end{eqnarray*}}
\def\LL{\left}
\def\RR{\right}
\def\nn{\nonumber\\}

\def\tr{\mathrm{tr}}

\def\bpsi{\bar\psi}
\def\gm5{\gamma_5}

\def\sign{\mathrm{sign}}

\def\bpsi{\bar{\psi}}
\def\ss#1{\slash\hspace{-2mm}#1}
\def\dl{\partial_\lambda D(p) }
\def\dli{\partial_\lambda D^{-1} (p) }
\def\ds{\partial_\sigma D(p) }
\def\dsi{\partial_\sigma D^{-1} (p) }
\def\dm{\partial_\mu D(p) }
\def\dmi{\partial_\mu D^{-1} (p) }
\def\dn{\partial_\nu D(p) }
\def\dni{\partial_\nu D^{-1} (p) }
\begin{document}
%


\begin{frontmatter}

\begin{flushright}
{\normalsize NTUTH-98-098}\\
\end{flushright}

\title{
Perturbation calculation of the axial anomaly
of Ginsparg-Wilson fermion }
\author{Ting-Wai Chiu and Tung-Han Hsieh}
\address{Department of Physics, National Taiwan University,
	Taipei, Taiwan 106, R.O.C}

\begin{abstract}
\noindent

We evaluate the axial anomaly for the general Ginsparg-Wilson fermion
operator $ D = D_c ( \Id + R D_c )^{-1} $ with $ R = r \Id $.
For any chirally symmetric $ D_c $ which in the free fermion limit
is free of species doubling and behaves like
$ i \gamma_\mu p_\mu $ as $ p \to 0 $, the axial anomaly
$ \tr [ \gamma_5 (R D) (x,x) ] $ for $ U(1) $ lattice gauge theory with
single fermion flavor is equal to
$ \frac{e^2}{32 \pi^2} \epsilon_{\mu\nu\lambda\sigma}
  F_{\mu\nu}(x) F_{\lambda\sigma}(x+\hat\mu+\hat\nu) $
up to higher order terms and/or non-perturbative contributions.
The $ F \tilde{F} $ term is $r$-invariant and has the correct
continuum limit.

\end{abstract}
\end{frontmatter}

\noindent
{\bf PACS \#:}  11.15.Ha, 11.30.Fs, 11.30.Rd 


\section{Introduction}
\noindent

In 1981, Ginsparg and Wilson \cite{gwr} formulated a criterion for
breaking the chiral symmetry of the massless Dirac operator on the
lattice,
\bea
\label{eq:gwr}
D \gamma_5 + \gamma_5 D = 2 a D \gamma_5 R D
\eea
where $ R $ is any invertible hermitian operator which is
local in the position space and trivial in the Dirac space, and $ a $
is the lattice spacing. The underlying reason for breaking the
continuum chiral symmetry on the lattice is due to the Nielson-Ninomiya
no-go theorem \cite{no-go} which states that any Dirac operator on the
lattice cannot simultaneously possess locality, free of species doubling
and the chiral symmetry. However, one prefers to violate the chiral symmetry
rather than the other two basic properties, is due to the fact that if the
chiral symmetry breaking is specified by the RHS of (\ref{eq:gwr})
having one $ \gamma_5 $ sandwiched by two Dirac operators, then
not only the usual chiral symmetry can be recovered in the continuum
limit $ a \to 0 $, but (\ref{eq:gwr}) also gaurantees the remnant
chiral symmetry on the lattice, that is, the action
$ {\mathcal A } = \bar \psi D \psi $
is invariant under the finite chiral transformation on the lattice
\bea
\label{eq:chi_ta}
\psi \rightarrow \exp [ \theta \gamma_5 ( \Id - a R D ) ] \psi \\
\label{eq:chi_tb}
\bar\psi \rightarrow \bar\psi \exp[ \theta ( \Id - a D R ) \gamma_5 ]
\eea
where $ \theta $ is a global parameter. The infinitesimal form of
(\ref{eq:chi_ta}) and (\ref{eq:chi_tb}) was first observed by
L\"uscher \cite{ml98:2}. In fact, the GW relation (\ref{eq:gwr}) can be
generalized to accommodate the asymmetric finite chiral transformations
on the lattice \cite{twc98:10b} having two different hermitian operators,
say $ S $ and $ T $ in (\ref{eq:chi_ta}) and (\ref{eq:chi_tb})
respectively, by replacing $ 2 R $ in (\ref{eq:gwr}) by $ S + T $.
Furthermore, the particular form of chiral symmetry
breaking on the RHS of (\ref{eq:gwr}) also implies the chiral Ward
identities, non-renormalization of vector and flavor non-singlet axial
vector currents, and non-mixing of operators in different chiral
representations \cite{ph98:2}.

In general, on the lattice, given any chirally symmetric Dirac operator
$ D_c $
\bea
\label{eq:Dc}
D_c \gamma_5 + \gamma_5 D_c = 0 \ ,
\eea
which is free of species doubling but nonlocal as a consequence of the
Nielson-Ninomiya no-go theorem, the general solution \cite{twc98:6a}
\bea
\label{eq:gen_sol}
D = D_c ( \Id + a R D_c )^{-1}
\eea
is a chiral symmetry breaking transformation which gives a class of Dirac
operators satisfying the GW relation (\ref{eq:gwr}). The general
solution (\ref{eq:gen_sol}) also implies that any zero mode of $ D $
is also a zero mode of $ D_c $, and vice versa. That is, the zero modes of
$ D $ are $R$-invariant. Then the index of $ D $ is also $R$-invariant
and is equal to the index of $ D_c $. Thus the chiral symmetry breaking
transformation (\ref{eq:gen_sol}) cannot generate a non-zero index
for $ D $ if the index of $ D_c $ is zero \cite{twc98:9a}.
It has become clear that a basic attribute of $ D ( D_c ) $ should be
introduced, and it is called topological characteristics in
ref. \cite{twc98:9a,twc98:10a}. In general, the topological characteristics
of a Dirac operator cannot be revealed by any perturbation calculations.
Therefore, if we obtain non-zero axial anomaly for $ D $ in a perturbation
calculation, it does not necessarily imply that the index of $ D $ must be
non-zero for topologically non-trivial background gauge fields.
The most reliable way to determine the topological characteristics
of a Dirac operator is to perform the following numerical test
\cite{twc98:4,twc98:10a}: turn on a topologically nontrivial smooth
background gauge field with constant field tensors, then check whether
there are any zero modes of $ D $, and measure the index of $ D $ versus
the topological charge of the background gauge field. It should be
emphasized that the topological characteristics of $ D $ is a basic
attribute of $ D $, which is not due to finite size effects,
but persists for any lattice sizes, at any lattice spacings and
in any background gauge configurations.

In Ginsparg and Wilson's original paper \cite{gwr}, the axial anomaly
for any $ D $ satisfying Eq. (\ref{eq:gwr}) was derived, and it
agrees with the continuum axial anomaly if $ D $ is free of species
doubling and in the free fermion limit behaves like
$ i \gamma_\mu p_\mu $ as $ p \to 0 $.
However, in their derivation, Eq. (\ref{eq:gwr}) is used to eliminate
$ R $ in some of the intermediate expressions, then at the final steps
of their computations, $ D $ is replaced by $ D_c $, which is equivalent
to setting $ R=0 $. Strictly speaking, their procedures are not completely
self-consistent since $ R $ should be kept nonzero throughout the entire
computation, otherwise Eq. (\ref{eq:gwr}) cannot be used to eliminate $ R $
from any expressions containing $ D \gamma_5 R D $.
In other words, setting $ R = 0 $ at their final steps of integrations
actually invalidates their previous steps of using Eq. (\ref{eq:gwr}) to
eliminate $ R $. The proper procedure should keep $ R \ne 0 $ throughout
the entire calculation, and then shows that the axial anomaly is independent
of $ R $, finally the limit $ R = 0 $ can be safely taken.
This motivated us to re-derive the axial anomaly
for a general $ D $ satisfying the Ginsparg-Wilson relation, with the
proper procedure, and in the context of recent developments.
Furthermore, the realization of the Atiyah-Singer index theorem
on a finite ( infinite ) lattice relies on the perturbation result
of the axial anomaly as well as the topological characteristics of $ D $.
This provided us further impetus to carry out the tedious computations
and present the details of our derivation in this paper.
We note in passing that in ref. \cite{kiku98:6,adams98:12} the chiral
anomaly for the overlap-Dirac operator \cite{hn97:7,hn98:1} is calculated,
and their results agree with the continuum axial anomaly.
However, the overlap-Dirac operator is only one of the solutions
satisfying the Ginsparg-Wilson relation, and of course their results
do not imply that the same axial anomaly will be obtained for a general
$ D $ with another $ D_c $.

If we also require $ D $ to satisfy the hermiticity condition
\bea
D^{\dagger} = \gamma_5 D \gamma_5
\eea
then $ D_c $ also satisfies this condition, and becomes an antihermitian
operator (~$ D_c^{\dagger} = - D_c $~)
which has one to one correspondence to a unitrary operator $ V $,
\beq
\label{eq:DcV}
D_c = \frac{ \Id + V }{ \Id - V }
\eeq
where $ V $ also satisfies the hermiticity condition
$ \gm5 V \gm5 = V^{\dagger} $.
Thus the inverse operator in Eq. (\ref{eq:gen_sol}) must exist, and the
general solution $ D $ is well defined.

In this paper, we evaluate the axial anomaly
$ \tr[ \gamma_5 ( R D )(x,x)] $
for $ R $ diagonal in the position space, i.e., $ R = r \Id $ with
parameter $ r $. Then from Eq. (\ref{eq:gen_sol}), we have
\bea
D = D_c ( \Id + r D_c )^{-1}
\label{eq:DDc}
\eea
We shall restrict our discussions to the $ U(1) $ gauge
theory with single fermion flavor. However, it is straightforward to
generalize our derivations to lattice QCD.
For any $ D_c $ which
in free fermion limit is free of species doubling and behaves like
$ i \gamma_\mu p_\mu $ as $ p \to 0 $, our perturbation calculation
shows that the $ F \tilde{F} $ term of the axial anomaly
$ \tr[ \gamma_5 ( R D ) (x,x)] $ is independent of $ r $ and
has the correct continuum limit, i.e.,
\bea
  r \ \tr [ \gamma_5 D(x,x) ]
= \frac{e^2}{ 32 \pi^2} \epsilon_{\mu\nu\lambda\sigma}
  F_{\mu\nu}(x) F_{\lambda\sigma}(x+\hat\mu+\hat\nu) + \mbox{ other terms }
\label{eq:anomaly}
\eea
where the field tensor on the lattice is defined as
\bea
\label{eq:Fuv}
F_{\mu\nu} (x) = \frac{1}{a}
[ A_\nu(x+\hat\mu) - A_\nu(x) - A_\mu(x+\hat\nu) + A_\mu(x) ]
\eea
The other terms in (\ref{eq:anomaly}) in principle cannot be computed
directly by any perturbation calculation to a finite order, however,
their sum over the entire lattice can be determined and has significant
impacts to the index theorem on the lattice, as we will show in section 2.

The outline of this paper is as follows. In section 2, we derive the
divergence of the axial vector current for the Dirac operator satisfying
the general Ginsparg-Wilson relation, and to set up the theoretical
framework for the perturbation calculation in section 3. The topological
characteristics of $ D $ is shown to emerge naturally as an integer
functional of $ D $, after the axial anomaly is summed over the entire
lattice. In section 3, we perform the derivation of the axial anomaly.
In section 4, we conclude and discuss.
In appendix A, we derive an identity for the $ F \tilde F $ terms.
In appendix B, we derive some useful properties of the kernel for
the vector current which are used in the derivation of axial anomaly
in section 3. In appendix C, we prove an identity for the Ginsparg-Wilson
kernel of the vector current.

\section{ The axial vector current and its divergence }

\noindent
In this section we derive the divergence of the axial vector current
$ J^{5}_{\mu}(x;A,D) $ for the Dirac operator $ D $ satisfying the
Ginsparg-Wilson relation of the general form \cite{twc98:10b}
\bea
\label{eq:gwr_gen}
D \gm5 ( \Id - S D ) + ( \Id - D T ) \gm5 D = 0
\eea
where $ S $ and $ T $ are arbitrary invertible hermitian operators which
are local in the position space and trivial in the Dirac space.
The action for exactly massless fermion in a background gauge field is
\bea
{\mathcal A} = \sum_{x, y} \bpsi_x D(x,y;A) \psi_y
\eea
where $ x $ and $ y $ are site indices, and the Dirac indices are
suppressed.
Then the action $ {\mathcal A } $ is invariant under the chiral
transformation
\bea
\label{eq:ct1}
\psi  \rightarrow \exp[ \theta \gm5 ( \Id - S D ) ] \psi \\
\label{eq:ct2}
\bpsi \rightarrow \bpsi \exp[ \theta ( \Id - D T ) \gm5 ]
\eea
where $ \theta $ is a global parameter.
Hence, the divergence of the ( associated Noether current )
axial vector current, $ \partial^{\mu} J^5_{\mu}(x) $,
can be extracted from the change of the action $ \delta {\mathcal A } $
under the infinitesimal local chiral transformation at the site $ x $,
\BA
\label{eq:LCT_a}
\psi_x \rightarrow \psi_x + \delta \theta_x \gm5 [ (\Id - S D ) \psi ]_x \\
\label{eq:LCT_b}
\bpsi_x \rightarrow \bpsi_x + \delta \theta_x [ \bpsi (\Id -  D T ) ]_x \gm5
\EA
with the prescription
\BA
\label{eq:noether}
{\mathcal A } \rightarrow {\mathcal A } +
              \delta \theta_x \partial^{\mu} J^5_{\mu}(x)
\EA
Then we obtain
\BA
\partial^{\mu} J^5_{\mu}(x) &=&    \bpsi_x \gm5 ( D \psi )_x
                                 + (\bpsi D)_x \gm5 \psi_x         \nn
                            & &  - ( \bpsi D T )_x \gm5 ( D \psi )_x
                                 - (\bpsi D)_x \gm5 ( S D \psi )_x
\label{eq:div5_n}
\EA
which satisfies the conservation law
\bea
\label{eq:Q5}
\sum_x \partial^{\mu} J^5_{\mu}(x) = 0
\eea
due to the exact chiral symmetry (\ref{eq:gwr_gen}) on the lattice.
Now we take the lattice to be finite and with periodic boundary
conditions, then we define $ \partial_\mu J_{\mu}^5 (x) $ by the
backward difference of the axial vector current
\BA
\partial_\mu J_{\mu}^5(x)=
\sum_{\mu} \ [ J_{\mu}^5 (x) - J_{\mu}^5 ( x - \hat{\mu} ) \ ]
\label{eq:divJ5_def}
\EA
such that it is parity even under the parity transformation, and
the conservation law Eq. (\ref{eq:Q5}) is also satisfied.

To evaluate the fermionic average of the divergence of the axial
vector current in a fixed background gauge field,
\BA
\label{eq:div5_f}
\LL< \partial^{\mu} J^5_{\mu}(x) \RR> &=&
\frac{1}{Z} \int [d\psi][d\bpsi] \partial^{\mu} J^5_{\mu}(x)
\exp( - \bpsi D \psi )  \\
Z &=& \int [d\psi][d\bpsi] \exp( - \bpsi D \psi )
\EA
one would encounter $ D^{-1} $ which is not well defined for the exactly
massless fermion in topologically nontrivial gauge background. Thus,
one needs to introduce an infinitesimal mass $ m $ which couples to the
chirally symmetric Dirac operator $ D_c $ in the following way
\cite{twc98:10b}
\BA
\hat D &=& (D_c + m) \left[ \Id + \frac{1}{2} ( S + T ) D_c \right]^{-1}
\label{eq:Dm}
\EA
and then evaluate the fermionic average
(\ref{eq:div5_f}) with $ D $ replaced by $ \hat D $, and finally take
the limit ( $ m \to 0 $ ), i.e.,
\BA
\label{eq:div5_fm}
\LL< \partial^{\mu} J^5_{\mu}(x) \RR> &=& \lim_{ m \to 0 }
\frac{1}{Z} \int [d\psi][d\bpsi] \partial^{\mu} J^5_{\mu}(x)
\exp( - \bpsi \hat D \psi )  \\
Z &=& \int [d\psi][d\bpsi] \exp( - \bpsi \hat D \psi )
\EA
Substituting $ \partial^{\mu} J^5_{\mu}(x) $ by Eq. (\ref{eq:div5_n}) and
using Eq. (\ref{eq:Dm}), we obtain
\BA
\LL< \partial^{\mu} J^5_{\mu}(x) \RR>
&=& - \tr [ ( \Id - DT )_{x x} \gm5 ] - \tr [ ( \Id - S D )_{x x} \gm5 ] \nn
&& + \ m \ \tr\LL( \LL\{ \left[ \Id-\frac{1}{2}(S+T)D \right]
                      \hat D^{-1}(\Id-DT) \RR\}_{x x} \gm5\RR) \nn
&& + \ m \ \tr\LL( \LL\{ (\Id-SD) \hat D^{-1}
                  \left[ \Id-\frac{1}{2}(S+T)D \right] \RR\}_{x x} \gm5 \RR)
\label{eq:div5_na}
\EA
where $ \tr $ denotes the trace in the Dirac space.
The first two terms on the RHS of (\ref{eq:div5_na}) is equal to
\BA
\tr\LL\{ \gm5 [ ( S + T ) D ]_{x x} \RR\}
\EA
while the last two terms in the limit ( $ m \to 0 $ ) give
\BA
\label{eq:zeromodes}
  2 \ m \ \tr [ \hat D^{-1}_{x x} \gm5  ]
\EA
which can be rewritten as
\BA
\label{eq:zeromodes_1}
  2 \ m \ \sum_{\alpha,\beta} \sum_{s}
\frac{ {\phi_s}^{\alpha}(x) \gm5^{\alpha \beta} [ {\phi_s}^{\beta} (x) ]^{*} }
     { m + \lambda_s }
\EA
where $ \phi_s $ and $ \lambda_s $ are normalized eigenfunction and
eigenvalue of $ D $,
\BA
\label{eq:eigen}
\sum_y \sum_{\beta} D_{xy}^{\alpha\beta} \phi_s^{\beta}(y)
= \lambda_s \phi_s^{\alpha}(x) \nn
\sum_x \sum_\alpha [\phi_s^{\alpha}(x)]^{*} \phi_{s'}^{\alpha}(x)
=\delta_{ss'}
\EA
In the limit ( $ m \to 0 $ ), only zero modes of $ D $ contribute to
(\ref{eq:zeromodes_1}) and the result is
\BA
     2 \sum_{s=1}^{N_{+}} [\phi_s^{+}(x)]^{\dagger} \phi_s^{+} (x)
   - 2 \sum_{t=1}^{N_{-}} [\phi_t^{-}(x)]^{\dagger} \phi_t^{-} (x)
\EA
where $ \phi_s^{+} $ and $ \phi_t^{-} $ are normalized eigenfunctions
of $ D $ ( $ D_c $ ) with eigenvalues $ \lambda_s = \lambda_t = 0 $ and
chiralities $ +1 $ and $ -1 $ respectively. Therefore in the
limit $ m \to 0 $, Eq. (\ref{eq:div5_na}) becomes
\BA
\LL< \partial^{\mu} J^5_{\mu}(x) \RR>
&=& \tr \LL\{ \gm5 [ ( S + T ) D ]_{x x} \RR\} \nn
& & + 2 \sum_{s=1}^{N_{+}} [\phi_s^{+}(x)]^{\dagger} \phi_s^{+} (x)
    - 2 \sum_{t=1}^{N_{-}} [\phi_t^{-}(x)]^{\dagger} \phi_t^{-} (x)
\label{eq:divJ5}
\EA
This is the anomaly equation for $ D $ satisfying the general
Ginsparg-Wilson relation (\ref{eq:gwr_gen}) which is
the exact chiral symmetry on the lattice, where the axial
vector current $ J_{\mu}^5(x) $ is the associated Noether current.

On the other hand, as usual, if one considers the action built from
the chirally symmetric part of $ D $ \cite{gwr},
\bea
\label{eq:A_s}
{\mathcal A}_s = \bpsi \ \frac{1}{2}( D - \gm5 D \gm5 ) \ \psi
\eea
then $ {\mathcal A}_s $ has the usual chiral symmetry, and the divergence
of the associated Noether current can be extracted from the change
of the action $ \delta {\mathcal A}_s $ under the infinitesimal local
chiral transformation at the site $ x $
\BAN
\psi_x  \rightarrow \psi_x  + \delta \theta_x \gm5    \psi_x \\
\bpsi_x \rightarrow \bpsi_x + \delta \theta_x \bpsi_x \gm5
\EAN
with the prescription (\ref{eq:noether}), and one obtains 
\BA
\partial^{\mu} J^5_{\mu}(x) &=&
  \frac{1}{2} ( \bpsi [ D, \gm5 ] )_x \psi_x
- \frac{1}{2} \bpsi_x ( [ D, \gm5 ] \psi )_x                     \nn
&=& \bpsi_x \gm5 ( D \psi )_x  + (\bpsi D)_x \gm5 \psi_x         \nn
&&  - \frac{1}{2} [ \bpsi D ( S + T ) \gm5 D ]_x \psi_x
    - \frac{1}{2} \bpsi_x [ D \gm5 ( S + T ) D \psi ]_x
\label{eq:div5_gw}
\EA
which also satisfies the conservation of total chiral charge,
Eq. (\ref{eq:Q5}), due to the usual chiral symmetry.
Although Eq. (\ref{eq:div5_gw}) looks fairly different from
Eq. (\ref{eq:div5_n}), however, the fermionic average of (\ref{eq:div5_gw})
in a background gauge field \cite{ph98:1} is equal to that of
(\ref{eq:div5_n}), i.e., Eq. (\ref{eq:divJ5}).
It is evident that the difference between (\ref{eq:div5_n}) and
(\ref{eq:div5_gw}) has no physical consequences.
In fact, it is possible to redefine
$ \partial^{\mu} J^5_{\mu}(x) $ of Eq. (\ref{eq:div5_n})
in many different ways provided that it satisfies the conservation law,
Eq. (\ref{eq:Q5}), and its fermionic average agrees with
Eq. (\ref{eq:divJ5}), however, the corresponding $ J_\mu^5(x) $ is not
equal to the Noether current associated to the exact
chiral symmetry on the lattice (\ref{eq:gwr_gen}).
For example, if one redefines (\ref{eq:div5_n}) as
\BA
\partial^{\mu} {\mathcal J}^5_{\mu}(x)
&=& \bpsi_x \gm5 ( D \psi )_x +  (\bpsi D)_x \gm5 \psi_x         \nn
&&  - \bpsi_x ( DT \gm5 D \psi )_x - (\bpsi D \gm5 SD )_x \psi_x  \nn
&=& [ \bpsi D \gm5 ( \Id - SD ) ]_x \psi_x
    + \bpsi_x [ ( \Id - DT) \gm5 D \psi ]_x  \nn
&=& [ \bpsi D \gm5 ( \Id - SD ) ]_x \psi_x
    - \bpsi_x [  D \gm5 ( \Id - SD ) \psi ]_x
\label{eq:div5_nc}
\EA
where Eq. (\ref{eq:gwr_gen}) has been used in the last equality, then
the fermionic average of Eq. (\ref{eq:div5_nc}) obviously agrees with
Eq. (\ref{eq:divJ5}), and the conservation law, Eq. (\ref{eq:Q5}) is
also satisfied.

Likewise, the divergence of the vector current can be extracted from the
change of the action $ \delta {\mathcal A } $ under the infinitesimal local
transformation
\BAN
\psi_x  \to \psi_x  + \delta\theta_x \psi_x  \\
\bpsi_x \to \bpsi_x - \delta\theta_x \bpsi_x
\EAN
with the prescription
\BAN
{\mathcal A } \to {\mathcal A } + \delta\theta_{x} \partial^{\mu} J_{\mu} (x),
\EAN
then we obtain
\BA
\partial_\mu J_{\mu} (x) = ( \bpsi D )_x \psi_x - \bpsi_x  ( D \psi )_x
\label{eq:divJ}
\EA
where $ \partial_\mu J_{\mu} (x) $ is defined by the backward difference
of the vector current
\BA
\partial_\mu J_{\mu}(x)=
\sum_{\mu} \ [ J_{\mu} (x) - J_{\mu} ( x - \hat{\mu} ) \ ]
\EA
such that it is parity even under the parity transformation,
and the conservation law due to the $ U_{V}(1) $ symmetry,
\bea
\label{eq:Qv}
\sum_x \partial_\mu J_{\mu}(x) = 0
\eea
is satisfied on a finite lattice with periodic boundary conditions.
If the vector current is expressed in terms of the kernel $ K_\mu $
as
\bea
\label{eq:ju}
J_{\mu}(x;A,D) = \sum_{y,z} \bpsi_{x+y} K_{\mu}(x,y,z;A,D) \psi_{x+z}
\eea
then by comparing Eqs. (\ref{eq:div5_gw}) and (\ref{eq:divJ}),
the axial vector current satisfying (\ref{eq:div5_gw}) can be written as
\bea
\label{eq:ju5}
J^{5}_{\mu}(x;A,D) = \sum_{y,z} \bpsi_{x+y} K^5_{\mu}(x,y,z;A,D) \psi_{x+z}
\eea
where the kernel $ K^5_{\mu} $ is related to $ K_\mu $ by
\bea
K^5_{\mu} = \frac{1}{2} ( K_{\mu} \gamma_5 - \gamma_5 K_{\mu} )
\label{eq:K5_gw}
\eea
%
%
We note in passing that for the $ {\mathcal J}_{\mu}^5 $
defined in (\ref{eq:div5_nc}), one could not find a simple relationship
between the kernel $ K_\mu^5 $ of this axial vector current and the
kernal $ K_\mu $ of the vector current, since
(\ref{eq:div5_nc}) is non-linear in $ D $.

Now summing Eq. (\ref{eq:divJ5}) over all sites of the lattice, the LHS
is zero due to the conservation law (\ref{eq:Q5}), then the RHS
gives the so called index theorem on the lattice \cite{ph98:1,ml98:2,hn97:7}
\BA
N_{-} - N_{+} = \frac{1}{2} \sum_x \tr\LL\{ \gm5 [ ( S + T ) D ]_{x x} \RR\}
\label{eq:index_thm}
\EA
where $ N_{+} ( N_{-} ) $ denotes the number of zero modes of positive
( negative ) chirality. It has been shown in ref. \cite{twc98:9a} that
the index is invariant with respect to $ S $ and $ T $ and is equal to
that at the chiral limit,
\BA
 N_{-} - N_{+} = \lim_{S, T \to 0 }
\frac{1}{2} \sum_x \tr\LL\{ \gm5 [ ( S + T ) D ]_{x,x} \RR\}
= \frac{1}{2} \sum_{x} \tr [ \gamma_5 V_{x,x} ]
\label{eq:ACV}
\EA
where Eq. (\ref{eq:DcV}) has been used. We note that {\em a priori},
there is no compelling reasons to gaurantee that $ D $ has zero modes
in topologically non-trivial sectors. It could happen that $ D $
is local and free of species doubling in the free fermion limit,
but turns out to have zero index in any background gauge fields
\cite{twc98:10a}.
In that case, $ D $ is called topologically trivial, and the index
theorem (\ref{eq:index_thm}) is trivially satisfied with both sides equal
to zero, however, it does not correspond to the Atiyah-Singer index
theorem in continuum. Presumably the index of $ D $ is a topological and
non-perturbative quantity, therefore the topological characteristics of
$ D $ cannot be revealed by any perturbation calculculations at finite
orders. We refer to ref. \cite{twc98:9a,twc98:10a} for further discussions
on topological characteristics of $ D $.
As we will show later in this section, the topological characteristics
of $ D $ emerges naturally as an integer functional of $ D $,
after the axial anomaly is summed over all lattice sites.

In the next section, we will show that the first term on the RHS of
Eq. (\ref{eq:divJ5}),
\beq
\label{eq:I2_ST}
\tr \{ \gamma_5 [ ( S + T ) D ]_{x,x} \}
\eeq
reproduces the topological charge density
\beq
\label{eq:ax_cont}
\rho(x) \equiv \frac{e^2}{16\pi^2} \epsilon_{\mu\nu\lambda\sigma}
               F_{\mu\nu}(x) F_{\lambda\sigma}(x+\hat\mu+\hat\nu)
\eeq
up to higher order terms and/or non-perturbative contributions.
The field tensor $ F_{\mu\nu}(x) $ on the
lattice is defined in (\ref{eq:Fuv}) for QED.
We note that in (\ref{eq:ax_cont}) the positions of the field tensors
$ F_{\mu\nu} $ and $ F_{\lambda\sigma} $ are chosen at $ x $ and
$ x + \hat\mu + \hat\nu $ respectively such that $ \rho(x) $ satisfies
\beq
\sum_x \delta_x \rho(x) = 0
\label{eq:delta_rho}
\eeq
for any local deformations of the gauge field \cite{ml98:8}.
In the continuum limit, $ \rho(x) $ agrees with the Chern-Pontryagin
density in continuum.

In order to extract the term which is quadratic in gauge fields
from (\ref{eq:I2_ST}), we consider the following operator
$ {\mathcal H} $ \cite{ph98:1} which is defined as
\BA
\label{eq:HO}
{\mathcal H} [ O ] \equiv \sum_{x, y} x_{\sigma}  y_{\lambda}
    {\delta\over\delta A_{\nu}(0)} {\delta\over\delta A_{\mu}(y)}
    \Bigl[ O(x) \Bigr] \Bigr|_{A=0} \
\EA
where $ O(x) $ is any observable. It is evident that the operator
$ {\mathcal H} $ picks up the terms quadratic in gauge fields
from the observable $ O $ and converts them into a constant.

The operator $ {\mathcal H} $ acting on (\ref{eq:ax_cont}) gives
\bea
{\mathcal H} [ \rho ] &=& \frac{e^2}{16\pi^2} \sum_{x, y}
    x_{\sigma} y_{\lambda}
    {\delta\over\delta A_{\nu}(0)} {\delta\over\delta A_{\mu}(y)}
    \Bigl[ \epsilon_{\alpha\beta\gamma\delta}
     F_{\alpha\beta}(x)
     F_{\gamma\delta}(x+\hat\alpha+\hat\beta)\Bigr] \Bigr |_{A=0} \\
&=& \frac{e^2}{ 2 \pi^2} \epsilon_{\mu\nu\lambda\sigma}
\label{eq:H_rho}
\eea
where Eq. (\ref{eq:Fuv}) has been used.
We note that it is not necessary to take the limit $ A_\mu = 0 $ after
the differentiations since $ \rho $ is quadratic in gauge fields.
The proof of Eq. (\ref{eq:H_rho}) is given in appendix A.
We note in passing that
$$
 {\mathcal H'} [ \epsilon_{\alpha\beta\gamma\delta} F_{\alpha\beta} (x)
                     F_{\gamma\delta}(x+\hat\alpha+\hat\beta) ]
= {\mathcal H'} [ \epsilon_{\alpha\beta\gamma\delta} F_{\alpha\beta} (x)
                     F_{\gamma\delta}(x) ]
= 8 \epsilon_{\mu\nu\lambda\sigma}
$$
where the operator $ {\mathcal H'} $ is similar to $ {\mathcal H } $
defined in (\ref{eq:HO}) but without imposing $ A_\mu = 0 $ after
differentiations with respect to the gauge fields. However we exclude
the case of $ \rho $ having both field tensors located at $ x $
since it does not satisfy Eq. (\ref{eq:delta_rho}).

On the other hand, if we have $ {\mathcal H} $ act on
(\ref{eq:I2_ST}) and obtain
\bea
{\mathcal H} \left( \tr \{ \gamma_5 [ ( S + T ) D ]_{x x} \} \right) =
\frac{e^2}{ 2 \pi^2} \epsilon_{\mu\nu\lambda\sigma}
\label{eq:ax_lat}
\EA
then we can infer that
\bea
\hspace{-5mm} \tr \{ \gamma_5 [ ( S + T ) D ]_{x x} \}
= \frac{e^2}{ 16 \pi^2} \epsilon_{\mu\nu\lambda\sigma}
            F_{\mu\nu}(x) F_{\lambda\sigma}(x+\hat\mu+\hat\nu)
  + \mbox{ other terms }
\label{eq:PT}
\eea
where "other terms" denotes those terms which cannot be determined
by the second order perturbation calculation using $ {\mathcal H} $,
which in general consists of higher order terms in $ A_\mu $ and/or
derivatives, plus terms due to
non-perturbative and/or topologcial effects (if any).
The other terms in (\ref{eq:PT}) in general cannot be computed
directly by any perturbation calculation to a finite order, however,
their sum over the entire lattice can be determined and might have
significant impacts to the index theorem on the lattice,
as we will show later.

Although Eq. (\ref{eq:PT}) refers to the infinite lattice, however,
we expect that it also holds for finite lattices. The argument
\cite{twc98:9a} is as follows.
If $ D $ is local, the boundary effects enter as
$ \sim \exp( - m(R) L / a ) $, where $ L $ is the lattice
size, $ a $ is the lattice spacing, $ m(R) $ is a monotonic
increasing function of $ R= r \Id $ ( take the simplest case ) with
$ m(0) $ equal to zero ( this is equivalent to that $ D_c $ is nonlocal )
and $ m(\infty) $ a positive constant.
As $ L \to \infty $, the finite size effects vanish and the axial
anomaly is given by Eq. (\ref{eq:PT}). For $ L $ is finite, one might
naively expect that the $ L $ dependence would enter the
$ F \tilde{F} $ term and "other terms", through the variable $ m(R) L / a $.
However, the $ F \tilde{F} $ term cannot have $ R $ dependence,
otherwise it would be in contrary to the fact that the index of $ D $ is
$R$-invariant \cite{twc98:9a}. Consequently, if $ L $ is gradually decreased
from infinity toward a finite value, all dependence of $ L $ and $ R $ only
resides in "other terms". So, Eq. (\ref{eq:PT}) also holds for finite
lattices provided that $ D $ is local.

After summing Eq. (\ref{eq:PT}) over all sites of the lattice, we have
\bea
& & 2 ( N_{-} - N_{+} ) = \sum_x \tr\LL\{\gm5 [(S+T)D]_{x,x} \RR\} \nn
&=& \sum_{x} \frac{e^2}{ 16 \pi^2} \epsilon_{\mu\nu\lambda\sigma}
               F_{\mu\nu}(x) F_{\lambda\sigma}(x+\hat\mu+\hat\nu)
    + \sum_x ( \mbox{ other terms } )
\label{eq:PT_sum}
\eea
Now we consider the topologically nontrivial background $ U(1) $ gauge field
on a 4-dimensional torus ( $ x_{\mu} \in [0,L_{\mu}], \mu = 1, \cdots, 4 $ ),
\bea
\label{eq:A1}
e A_1(x) &=& \frac{ 2 \pi h_1 }{L_1} - \frac{ 2 \pi q_1 x_2 }{ L_1 L_2 }
         +  A_1^{(0)} \sin \left( \frac{ 2 \pi n_2 }{L_2} x_2 \right)  \\
\label{eq:A2}
e A_2(x) &=& \frac{ 2 \pi h_2 }{L_2}
         +  A_2^{(0)} \sin \left( \frac{ 2 \pi n_1 }{L_1} x_1 \right)  \\
\label{eq:A3}
e A_3(x) &=& \frac{ 2 \pi h_3 }{L_3} - \frac{ 2 \pi q_2 x_4 }{ L_3 L_4 }
         +  A_3^{(0)} \sin \left( \frac{ 2 \pi n_4 }{L_4} x_4 \right)  \\
\label{eq:A4}
e A_4(x) &=& \frac{ 2 \pi h_4 }{L_4}
         +  A_4^{(0)} \sin \left( \frac{ 2 \pi n_3 }{L_3} x_3 \right)
\eea
where $ q_1 $, $ q_2 $, $ n_1, \cdots,  n_4 $ are integers.
The global part is characterized by the topological charge
\beq
Q = \frac{e^2}{32\pi^2} \int d^4 x \ \epsilon_{\mu\nu\lambda\sigma} \
     F_{\mu\nu}(x) F_{\lambda\sigma}(x)  = q_1 q_2
\label{eq:ntop}
\eeq
which must be an integer. The harmonic parts are parameterized by four
real constants $ h_1 $, $ h_2 $, $ h_3 $ and $ h_4 $.
The local parts are chosen to be sinusoidal fluctuations with
amplitudes $ A_1^{(0)} $, $ A_2^{(0)} $, $ A_3^{(0)} $ and $ A_4^{(0)} $,
and frequencies $ \frac{ 2 \pi n_2 }{L_2} $, $ \frac{ 2 \pi n_1 }{L_1} $,
$ \frac{ 2 \pi n_4 }{L_4} $ and $ \frac{ 2 \pi n_3 }{L_3} $ respectively.
The discontinuity of $ A_1(x) $ ( $ A_3(x) $ ) at $ x_2 = L_2 $
( $ x_4 = L_4 $ ) due to the global part only amounts to a gauge
transformation. The field tensors $ F_{12} $ and $ F_{34} $ are
continuous on the torus, while other $ F's $ are zero.
To transcribe the continuum gauge fields to link variables on a finite
lattice with periodic boundary conditions,
we take the lattice sites at $ x_\mu = 0, a, ..., ( N_\mu - 1 ) a $,
where $ a $ is the lattice spacing and $ L_\mu = N_\mu a $ is the lattice
size. Then the link variables are
\bea
\label{eq:U1}
U_1(x) &=& \exp \left[ \text{i} e A_1(x) a \right] \\
\label{eq:U2}
U_2(x) &=& \exp \left[ \text{i} e A_2(x) a
 + \text{i} \delta_{x_2,(N_2 - 1)a} \frac{ 2 \pi q_1 x_1 }{L_1} \right] \\
\label{eq:U3}
U_3(x) &=& \exp \left[ \text{i} e A_3(x) a \right] \\
\label{eq:U4}
U_4(x) &=& \exp \left[ \text{i} e A_4(x) a
 + \text{i} \delta_{x_4,(N_4 - 1)a} \frac{ 2 \pi q_2 x_3 }{L_3} \right]
\eea
The last term in the exponent of $ U_2(x) $ ( $ U_4(x) $ ) is included to
ensure that the field tensor $ F_{12} $ ( $ F_{34} $ ) which is defined by
the ordered product of link variables around a plaquette
[ Eq. (\ref{eq:Fuv_b}) ] is continuous on the torus.
Then the topological charge of this gauge configuration on the finite
lattice is
\bea
Q = \frac{e^2}{32 \pi^2} \sum_x \ \epsilon_{\mu\nu\lambda\sigma}
    F_{\mu\nu}(x) F_{\lambda\sigma}(x+\hat\mu+\hat\nu) = q_1 q_2
\label{eq:ntop_lat}
\eea
which agrees with the topological charge [ Eq. (\ref{eq:ntop}) ]
on the 4-dimensional torus. From Eq. (\ref{eq:PT_sum}), we obtain
\bea
\sum_x ( \mbox{ other terms } ) = 2 ( N_{-} - N_{+} - q_1 q_2 )
\label{eq:higher}
\eea
which is also an integer. Since $ D $ does not have any genuine zeromodes
in the topologically trivial gauge background,
thus the integer (\ref{eq:higher}) must be proportional to $ q_1 q_2 $
and can be represented as
\bea
& & \sum_x ( \mbox{ other terms } ) \nn
&=& ( c[D] - 1 ) \sum_{x} \frac{e^2}{ 16 \pi^2}
                          \epsilon_{\mu\nu\lambda\sigma}
     F_{\mu\nu}(x) F_{\lambda\sigma}(x+\hat\mu+\hat\nu) \nonumber
\eea
where $ c[D] $ is an integer functional of $ D $.
Then Eq. (\ref{eq:PT_sum}) becomes
\bea
 N_{-} - N_{+}
&=& \frac{1}{2} \sum_x \tr\LL\{ \gm5 [ ( S + T ) D ]_{x x} \RR\} \nn
&=& c[D] \sum_{x} \frac{e^2}{ 32 \pi^2} \epsilon_{\mu\nu\lambda\sigma}
            F_{\mu\nu}(x) F_{\lambda\sigma}(x+\hat\mu+\hat\nu) \nn
&=& c[D] \ Q
\label{eq:ASL}
\eea
where $ c[D] = c[D_c] $ due to the invariance of the zero modes and
the index under the chiral symmetry breaking transformation
(\ref{eq:gen_sol}) \cite{twc98:9a}.
In general, Eq. (\ref{eq:ASL}) holds for any smooth background gauge
configurations on a finite ( infinite ) lattice provided that the
topological charge on the lattice (\ref{eq:ntop_lat}) is an integer
and the axial anomaly on the lattice satisfies Eq. (\ref{eq:ax_lat}).
It is remarkable that, if any of these higher order,
non-perturbative and topological contributions to the axial anomaly
does exist, then their total effects to the index can only enter as an
integer ( $ c[D] - 1 $ ) multiple of the topological charge of the
background gauge field. We emphasize that the emergence of $ c[D] $
is not due to the finite size of the lattice but an intrinsic
characteristics of $ D $, in particular, when $ c[D] = 0 $, then
the axial anomaly vanishes at each site of the lattice, independent of
the size of the lattice. In general, the topological
characteristics, $ c[D] $, is an integer functional of $ D $, which in
turn depends on some parameters of $ D $ as well as the gauge configuration.
$ c[D] $ could become chaotic when the background gauge field is rough,
as first demonstrated in ref. \cite{twc98:10a}.
For smooth gauge configurations, $ D $ can be classified
according to its topological characteristics as follows \cite{twc98:9a}.
If $ c[D] = 1 $, then $ D $ is called {\em topologically proper};
else if $ c[D] = 0 $, then $ D $ is called {\em topologically trivial};
else $ c[D] = \mbox{integer} \ne 0,1 $, then $ D $ is called
{\em topologically improper}. {\em Only for $ D $ is topologically proper,
i.e., $ c[D] = 1 $, Eq. (\ref{eq:ASL}) can realize the Atiyah-Singer index
theorem on a finite ( infinite ) lattice.}
[ Note the invariance of the topological
charge in Eqs. (\ref{eq:ntop_lat}) and (\ref{eq:ntop}) ].
Eq. (\ref{eq:ASL}) provides the theoretical understanding of the
numerical results \cite{twc98:4,twc98:10a} for exact zero modes
satisfying the Atiyah-Singer index theorem even on very small lattices
in two dimensions as well as in four dimensions.

We note that in ref. \cite{twc98:10a}, using the exact reflection
symmetry and the exact solution of the free fermion propagator,
the ( lowest order ) perturbation theory is shown to break down
at the topological phase boundaries in the $ m_0 $ parameter space
of the overlap-Dirac operator.
The zero index of $ D $ at the phase boundaries can be interpreted
as $ c[D]=0 $, due to non-perturbative and/or topological contributions.
In general, $ c[D] $ incorporates all kinds of contributions from
all fermionic modes.

We also note that in the operator $ {\mathcal H} $,
the gauge fields are set to zero after the
differentiations with respect to the gauge fields.
This implies that {\it only free fermion propagators}
are needed in the evaluation of the LHS of Eq. (\ref{eq:ax_lat}).
In the next section, we will show that Eq. (\ref{eq:ax_lat}) is
indeed satisfied by the Ginsparg-Wilson Dirac operator (\ref{eq:gen_sol})
with $ R = r \Id $, provided that $ D_c $ in the free fermion limit is
free of species doubling and behaves like
$ i \gamma_\mu p_\mu $ as $ p \to 0 $.

Recently L\"uscher \cite{ml98:8} proved that for $ U(1) $ lattice
gauge theory, and for $ S + T = 2 $, if the axial anomaly
$ q(x;A,D)$ $ = \tr [ \gamma_5 D_{x,x} ] $
satisfies
\bea
\sum_x \delta_x q(x;A,D) = 0
\label{eq:local}
\eea
for any local deformations of the gauge field, then
\bea
q(x;A,D) = \gamma \ \epsilon_{\mu\nu\lambda\sigma} F_{\mu\nu} (x)
                                    F_{\lambda\sigma}(x+\hat\mu+\hat\nu)
                 + \partial_\mu G_\mu(x;A,D)
\label{eq:qFkR}
\eea
where $ \gamma $ is a constant times an integer factor, and
\bea
\partial_\mu G_\mu(x;A,D) =
\sum_\mu \left[ G_\mu(x;A,D) - G_\mu(x-\hat\mu;A,D) \right].
\eea
The explicit form of the current $ G_\mu(x;A,D) $ is supposed to be very
complicated. As discussed in ref. \cite{twc98:9a},
Eq. (\ref{eq:qFkR}) can be generalized to any $ D $ satisfying
(\ref{eq:gwr_gen}) with all the $S$ and $T$ dependences residing in the
term $ \partial_\mu G_\mu(x;A,D) $, while the $ F \tilde{F} $ term
depends on the topological characteristics $ c[D] $ which is
$(S,T)$-invariant ( $ c[D] = c[D_c] $ ), i.e.,
\bea
\label{eq:qST}
q(x;A,D) &=& \frac{1}{2} \tr \{ \gamma_5 [( S + T ) D ]_{x,x} \}  \\
         &=& \gamma^{'} \ c[D_c] \epsilon_{\mu\nu\lambda\sigma} F_{\mu\nu} (x)
                                    F_{\lambda\sigma}(x+\hat\mu+\hat\nu)
                 + \partial_\mu G_\mu(x;A,D)
\label{eq:qFST}
\eea
where the current $ G_\mu(x;A,D) $ in general is a functional of $S$ and
$T$. Although Eq. (\ref{eq:qFkR}) is proved for the infinite lattice
in ref. \cite{ml98:8}, however, as discussed in ref. \cite{twc98:9a},
if $ D $ is local, then the boundary effects which depend on $ (S,T) $ and
the lattice size $ L $, can only enter the term $ \partial_\mu G_\mu(x;A,D) $,
thus Eqs. (\ref{eq:qFkR}) and (\ref{eq:qFST}) are also true for finite
lattices. The same argument has been presented in details to assert that
Eq. (\ref{eq:PT}) also holds for finite lattices.
[ See the second paragraph after Eq. (\ref{eq:PT}). ]

Now applying the operator $ {\mathcal H} $ to Eq. (\ref{eq:qFST}) and using
Eq. (\ref{eq:H_rho}), we obtain
\bea
{\mathcal H} \left[ q(x;A,D) \right] &=&
\gamma^{'} \ {\mathcal H} \left[ c[D] \
                  \epsilon_{\alpha\beta\gamma\delta} F_{\alpha\beta} (x)
                     F_{\gamma\delta}(x+\hat\alpha+\hat\beta) \right]
\nn
&&
                 + {\mathcal H} [ \partial_\alpha G_\alpha (x;A,D) ] \nn
&=& 8 \gamma^{'} \ \epsilon_{\mu\nu\lambda\sigma}
    + {\mathcal H} [ \partial_\alpha G_\alpha (x;A,D) ]
\label{eq:ANL}
\eea
where $ c[D] $ is presumably non-perturbative and thus cannot be determined
by the second order operator $ {\mathcal H} $ acting
on $ q(x;A,D) $. Hence $ D $ can only be assumed to be topologically
proper and $ c[D] $ is set to unity throughout the entire perturbation
calculation. Then the assertion of Eq. (\ref{eq:ax_lat}) implies that the
LHS of Eq. (\ref{eq:ANL}) is $(S,T)$-invariant. On the other hand, on
the RHS of (\ref{eq:ANL}), the first term is $(S,T)$-invariant, but
the second term in general depends on $S$ and $T$. This implies that
$ {\mathcal H} [ \partial_\alpha G_\alpha(x;A,D) ] $ must vanish identically,
and Eq. (\ref{eq:ANL}) gives
\bea
\gamma^{'} = \frac{e^2}{ 32 \pi^2 }
\eea
and (\ref{eq:qFST}) becomes
\bea
q(x;A,D) = \frac{e^2}{ 32 \pi^2 } \ c[D_c] \epsilon_{\mu\nu\lambda\sigma}
            F_{\mu\nu} (x) F_{\lambda\sigma}(x+\hat\mu+\hat\nu)
                                    + \partial_\mu G_\mu(x;A,D)
\label{eq:qFST_c}
\eea
For the $ U(1) $ gauge fields defined in Eqs. (\ref{eq:U1})-(\ref{eq:U4}),
summing Eq. (\ref{eq:qFST_c}) over all lattice sites yields
$$
N_{-} - N_{+} = c[D] \ Q
$$
in agreement with Eq. (\ref{eq:ASL}).

It is interesting to note that the integer factor in the $ \gamma $ of
Eq. (\ref{eq:qFkR}) turns out can be used to accomodate non-perturbative
and/or topological effects, and it is identified to be the topological
characteristics, $ c[D] $ which is first introduced in
ref. \cite{twc98:9a,twc98:10a} in the study of the index of $ D $
with repsect to the background gauge fields.
It also provides plausible explanations to some seemingly paradoxial
situations that one obtains the correct axial anomaly in the perturbation
calculation for a given $ D $ but the numerical ( nonperturbative )
computations give exactly zero axial anomaly at each site, for any lattice
sizes and for any gauge configurations.



\section{The Axial Anomaly}

In this section, we assert Eq. (\ref{eq:ax_lat}) for $ S + T = 2 r \Id $
by evaluating
\BA
\hspace{-5mm}
I_2 & \equiv & {\mathcal H} [ 2 \ r \ \tr( \gm5 D_{x,x} )  ]    \nn
    & = & - 2 \ r \ \sum_{m,n} n_{\sigma} m_{\lambda} \LL.\LL\{
    {\delta\over\delta A_{\nu}(0)}{\delta\over\delta A_{\mu}(m)}
        \LL< (\bpsi D \gamma_5 D)_n\psi_n \RR> \RR\}
    \RR|_{A=0}
\label{eq:I2}
\EA
where $ D $ satisfies the Ginsparg-Wilson
relation (\ref{eq:gwr}) with $ R = r \Id $, i.e.,
\bea
\label{eq:Dcr}
D = D_c ( \Id + r D_c )^{-1}
\eea
for any chirally symmetric Dirac operator $ D_c $ which
in the free fermion limit, is free of species doubling and
behaves like $ \ i \gamma_\mu p_\mu $ as $ p \to 0 $.

In general, we can write $ D_c $ in the momentum space as
\bea
\label{eq:Dcc}
D_c(p) = i \gamma_{\mu} C_{\mu}(p)
\eea
where $ C_\mu(p) $ are arbitrary functions which satisfy the following
properties in the free fermion limit : \\
(i)  $ C_\mu(p) \to p_\mu $ as $ p \to 0 $. \\
(ii) $ C_\mu(p) $ has no zeros in the entire Brillouin zone except
     at the origin $ p = 0 $. \\
Then using Eqs. (\ref{eq:Dcr}) and (\ref{eq:Dcc}), we obtain
\BA
\label{eq:D0}
   D &=& {i\ss{C} + rC^2\over 1 + r^2 C^2} \\
\label{eq:D0i}
    D^{-1} &=& {-i\ss{C}\over C^2} + r
\EA

First we perform the differentiations with respect to
the gauge fields. The general formula is
\BA
&& {\delta\over\delta A_{\nu}(0)}{\delta\over\delta A_{\mu}(m)} \LL<T(U)\RR>
\nn
&\equiv&
    {\delta\over\delta A_{\nu}(0)}{\delta\over\delta A_{\mu}(m)} \LL\{
        {1\over Z(U)}\int_{\bpsi\psi} T(U) e^{-\bpsi  D(U)\psi}
    \RR\}
\equiv
    \sum_{i=1}^{12} P_i
\EA
where $ U $ denotes link variables and $ T(U) $ is an arbitrary functional
of link variables and fermion fields, and
\BA
\label{eq:P1}
P_1 &=&
    \LL< T(U)
        \LL(\bpsi{\delta\over\delta A_{\mu}(m)} D(U)\psi\RR)
        \LL(\bpsi{\delta\over\delta A_{\nu}(0)} D(U)\psi\RR)
    \RR>
\\
P_2 &=&
    - \LL<T(U) 
	\LL(\bpsi{\delta\over\delta A_{\nu}(0)}
                  {\delta\over\delta A_{\mu}(m)} D(U)\psi\RR)
    \RR>
\\
P_3 &=&
    - \LL< \LL({\delta\over\delta A_{\nu}(0)}T(U)\RR)
        \LL(\bpsi{\delta\over\delta A_{\mu}(m)} D(U)\psi\RR)
    \RR>
\\
P_4 &=&
    - \LL< \LL({\delta\over\delta A_{\mu}(m)}T(U)\RR)
        \LL(\bpsi{\delta\over\delta A_{\nu}(0)} D(U)\psi\RR)
    \RR>
\\
P_5 &=&
    \LL<
	\LL({\delta\over\delta A_{\mu}(m)}{\delta\over\delta A_{\nu}(0)}
	T(U)\RR) 
    \RR>
\\
P_6 &=&
    -\LL< T(U)
        \LL(\bpsi{\delta\over\delta A_{\mu}(m)} D(U)\psi\RR)
    \RR>
    \LL< \LL(\bpsi{\delta\over\delta A_{\nu}(0)} D(U)\psi\RR) \RR>
\\
P_7 &=&
    \LL< \LL({\delta\over\delta A_{\mu}(m)}T(U)\RR) \RR>
    \LL< \LL(\bpsi{\delta\over\delta A_{\nu}(0)} D(U)\psi\RR) \RR>
\\
P_8 &=& 
    -\LL< T(U)\LL(\bpsi{\delta\over\delta A_{\nu}(0)} D(U)\psi\RR) \RR>
    \LL< \LL(\bpsi{\delta\over\delta A_{\mu}(m)} D(U)\psi\RR) \RR>
\\
\label{eq:P9}
P_9 &=&
    \LL< \LL({\delta\over\delta A_{\nu}(0)}T(U)\RR) \RR>
    \LL< \LL(\bpsi{\delta\over\delta A_{\mu}(m)} D(U)\psi\RR) \RR>
\\
P_{10} &=&
    - \LL< T(U) \RR>
    \LL< \LL(\bpsi{\delta\over\delta A_{\mu}(m)} D(U)\psi\RR)
                    \LL(\bpsi{\delta\over\delta A_{\nu}(0)} D(U)\psi\RR)
    \RR>
\\
P_{11} &=&
    \LL< T(U) \RR>
    \LL< \LL(\bpsi{\delta\over\delta A_{\nu}(0)}
        {\delta\over\delta A_{\mu}(m)} D(U)\psi\RR) \RR>
\\
\label{eq:P12}
P_{12} &=&
    2\LL< T(U) \RR>
    \LL< \LL(\bpsi{\delta\over\delta A_{\mu}(m)} D(U)\psi\RR) \RR>
    \LL< \LL(\bpsi{\delta\over\delta A_{\nu}(0)} D(U)\psi\RR) \RR>
\EA
To evaluate $ I_2 $ in (\ref{eq:I2}), we set
\BA
T(U) = (\bpsi D R \gm5 D)_n \psi_n =  r (\bpsi D \gm5 D)_n \psi_n
\label{eq:TU}
\EA
Then $ \LL<T(U)\RR> = - r \ \tr[ \gm5 D_{n,n} ] $ which vanishes in the
free fermion limit. So, $ P_{10} $, $ P_{11} $ and $ P_{12} $ do not
contribute to $ I_2 $.
Next we express the vector current in terms of the kernel $ K_{\mu} $
as
$$
J_{\mu}(k,U) = \sum_{i,j} \bpsi_{k+i} K_{\mu}(k,i,j;U) \psi_{k+j}.
\label{eq:JVU_def}
$$
However, the vector current $ J_{\mu} $ satisfying the divergence
condition Eq. (\ref{eq:divJ}) is not unique. A general and explicit
realization is given by Ginsparg and Wilson \cite{gwr} with
$ K_\mu(k,i,j;U) $ equal to $ D_{k+i,k+j} (U) $ times
the sign of $ (i-j)_\mu $ and times the fraction of the shortest
length paths from $k+j$ to $k+i$ which pass through the link from
$k$ to $k+\hat\mu$. The Ginsparg-Wilson kernel can be
expressed in terms of the derivative of the Dirac operator
with respect to the gauge field,
\BA
J_{\mu}(k,U) = \sum_{i,j} \bpsi_{k+i} K_{\mu}(k,i,j;U) \psi_{k+j}
= i \sum_{m,n} \bpsi_m {\delta\over\delta A_{\mu}(k)} D_{mn}(U) \psi_n
\label{eq:JVU}
\EA
where the second equality is proved in appendix C.
In the free fermion limit, the action is translational invariant,
$ D_{mn} = D_{m-n} $, then Eq. (\ref{eq:JVU}) becomes
\BA
J_{\mu}(k) = \sum_{i,j} \bpsi_{k+i} K_{\mu}(i,i-j) \psi_{k+i-j}
= \LL. i \sum_{m,n} \bpsi_m
      {\delta\over\delta A_{\mu}(k)}  D_{mn}(U) \psi_n \ \RR|_{A=0}
\label{eq:JV}
\EA
with
\BA
\label{eq:Kmu}
   K_{\mu}(i,i-j) = \sign(j_{\mu}) \ f_{\mu}(i,j) \ D_j
\EA
where $ f_{\mu}(i,j) $ is equal to the fraction of the shortest length
paths from $0$ to $j$ which pass through the link from $i-\hat\mu$ to $i$.

We note that Hasenfratz \cite{ph98:3} also showed that
$$
J_{\mu}(k,U) =
i \sum_{m,n} \bpsi_m {\delta\over\delta A_{\mu}(k)} D_{mn}(U) \psi_n
$$
satisfies the divergence condition, Eq. (\ref{eq:divJ}).
However, an explicit realization of the kernel has not been prescribed.
Furthermore, in the free fermion limit, the kernel is shown \cite{ph98:3}
to satisfy the so called sum rules which are equivalent to
Eqs. (\ref{eq:Id1}) and (\ref{eq:Id2}) proved in appendix B as well as
in ref. \cite{gwr}.

According to the properties of $ K_{\mu}(k,i,j;U) $ discussed
above and the following simple identity of the derivative
of a link variable with respect to the gauge field,
\BE
\label{eq:dUdA}
{\delta\over\delta A_{\nu}(n)} U_{\mu}(k) =
    i\delta_{\mu\nu}\delta_{kn} U_{\mu}(k)
\EE
we obtain
\BE
\label{eq:dKdA}
\LL.{\delta\over\delta A_{\nu}(n)} K_{\mu}(k,i,j;U)\RR|_{A=0} =
    c_{\mu\nu}(n) K_{\mu}(i,i-j)
\EE
where $ c_{\mu\nu}(n) $ are some constants independent of the gauge fields.
Some useful properties of $ K_\mu (i,i-j) $ are derived in Appendix B and
will be used in our evaluation of $ I_2 $.

Using Eqs. (\ref{eq:TU}), (\ref{eq:JVU}), (\ref{eq:JV}), and (\ref{eq:dKdA}),
the expressions of $P_i's$ in Eqs.(\ref{eq:P1})-(\ref{eq:P9})
become
\BA
P_1 &=& 
    -\LL< (\bpsi DR\gm5 D)_n\psi_n
	J_{\mu}(m)J_{\nu}(0) \RR>
\\
P_2 &=& ic_{\mu\nu}(0)
    \LL< (\bpsi DR\gm5 D)_n \psi_n J_{\mu}(m) \RR>
\\
P_3 &=& i
    \LL<\LL(\LL.{\delta\over\delta A_{\nu}(0)}
     (\bpsi DR\gm5 D)_n\psi_n
	\RR|_{A=0}\RR) J_{\mu}(m)\RR>
\\
P_4 &=& i
    \LL<\LL(\LL.{\delta\over\delta A_{\mu}(m)}
     (\bpsi DR\gm5 D)_n\psi_n \RR|_{A=0}\RR) J_{\nu}(0)\RR>
\\
P_5 &=& 
    \LL<\LL.{\delta\over\delta A_{\nu}(0)}{\delta\over\delta A_{\mu}(m)}
     (\bpsi DR\gm5 D)_n\psi_n \RR|_{A=0}\RR>
\\
\label{eq:P6'}
P_6 &=& 
    \LL< (\bpsi DR\gm5 D)_n\psi_n
	J_{\mu}(m)\RR> \LL<J_{\nu}(0)\RR>
\\
P_7 &=& -i
    \LL<\LL.{\delta\over\delta A_{\mu}(m)}
     (\bpsi DR\gm5 D)_n\psi_n \RR|_{A=0}\RR> \LL<J_{\nu}(0)\RR>
\\
P_8 &=& 
    \LL< (\bpsi DR\gm5 D)_n \psi_n J_{\nu}(0)\RR> \LL<J_{\mu}(m)\RR>
\\
\label{eq:P9'}
P_9 &=& -i
    \LL<\LL.{\delta\over\delta A_{\nu}(0)}
    (\bpsi DR\gm5 D)_n\psi_n \RR|_{A=0}\RR> \LL<J_{\mu}(m)\RR>
\EA
where the free fermion limit has been imposed for the
RHS of Eq. (\ref{eq:I2}), i.e.
\bea
\label{eq:I2'}
I_2 = - 2 \  \sum_{i=1}^{9} \Bigl\{ \sum_{m,n} n_{\sigma} m_{\lambda}
             P_i \Bigr\}
\eea
In above expressions, only $P_1$ has non-vanishing contributions to $I_2$.
This can be shown in the following. First consider the common factor
$ \LL<J_\nu (0)\RR> $ of $ P_6 $ and $ P_7 $. Using Eqs. (\ref{eq:JV})
and (\ref{eq:Kmu}), we obtain
\bea
\LL<J_{\nu}(0)\RR> &=& - \sum_{i,j} \tr [ D_{-j}^{-1} \ K_\nu(i,i-j) ] \nn
&=& \sum_{i,j} \tr [ D_{-j}^{-1} \ \sign(j_\nu) \ f(i,-j) \ D_{-j} ] \nn
&=& - \tr(\Id) \sum_{i,j}  \sign(j_\nu) \ f(i,j) = 0
\eea
since $ \sum_i f(i,j) = \sum_i f(i,-j ) $. Similarly, we can show that
$ \LL<J_\mu(m)\RR> = 0 $ since it is translational invariant.
Then $ P_6 $, $ P_7 $, $ P_8 $ and $ P_9 $ in
Eqs. (\ref{eq:P6'})-(\ref{eq:P9'}) are zero.

For $ P_2 $, the factor $ \LL< (\bpsi D R \gm5 D)_n \psi_n J_{\mu}(m) \RR> $
enters (\ref{eq:I2'}) to give
\BA
& & \sum_{m,n} n_{\sigma}  m_{\lambda}
    \LL<(\bpsi DR\gm5 D)_n\psi_n J_{\mu}(m)\RR>
\nn
&=&
    \sum_{m,n,i,j} n_{\sigma}  m_{\lambda}
    \LL< (\bpsi DR\gm5 D)_n \psi_n \bpsi_{m+i} K_{\mu}(i,i-j)\psi_{m+i-j}\RR>
\nn
&=&
    -r \sum_{m,n,i,j} n_{\sigma}  m_{\lambda}
       \Bigl\{  \tr[ \gm5 D_{m+i-j-n} D^{-1}_{n-m-i} K_{\mu}(i,i-j) ] \nn
& & \hspace{30mm} -\tr[ \gm5 D_0 ] \ \tr[ K_{\mu}(i,i-j) D^{-1}_{-j} ]
       \Bigr\}
\nn
&=& 0 \nonumber
\EA
where the identity
\bea
\label{eq:gm5}
\tr( \gm5 ) = \tr( \gm5 \gamma_\mu ) = \tr ( \gm5 \gamma_\mu \gamma_\nu ) =
\tr ( \gm5 \gamma_\mu \gamma_\nu \gamma_\sigma ) = 0
\eea
has been used, and the fact that
each factor of $ D $, $ D^{-1} $ or $ K_\mu $ can give at most
one gamma matrix in the free fermion limit [
Eqs. (\ref{eq:D0})-(\ref{eq:D0i}) and (\ref{eq:Kmu}) ].

For $ P_5 $, it enters (\ref{eq:I2'}) to give
\BA
& & \sum_{m,n} n_{\sigma} m_{\lambda}
    \LL<\LL.{\delta\over\delta A_{\nu}(0)}{\delta\over\delta A_{\mu}(m)}
     (\bpsi D R \gm5 D)_n \psi_n \RR|_{A=0} \RR> \nn
&=& \frac{1}{2} \sum_{m,n,i} n_{\sigma} m_{\lambda}
    \LL<\LL.{\delta\over\delta A_{\nu}(0)}{\delta\over\delta A_{\mu}(m)}
     \bpsi_i ( D_{i,n} \gm5 + \gm5 D_{i,n} ) \psi_n \RR|_{A=0} \RR>
\EA
where Eq. (\ref{eq:gwr}) has been used. Using Eq. (\ref{eq:dUdA}),
we immediately see that this expression only involves trace operations on
terms containing one $ \gm5 $ and less than four $ \gamma_{\mu}'s $ matrices,
thus it must be zero.

For $ P_3 $, it enters (\ref{eq:I2'}) to give
\bea
& & \sum_{m,n} n_{\sigma} m_{\lambda}
    \LL<\LL(\LL.{\delta\over\delta A_{\nu}(0)}
     (\bpsi DR\gm5 D)_n\psi_n
        \RR|_{A=0}\RR) J_{\mu}(m)\RR>                        \nn
&=& \frac{1}{2} \sum_{m,n,k} n_{\sigma} m_{\lambda}
    \LL<\LL(\LL.{\delta\over\delta A_{\nu}(0)}
      \bpsi_k [ D_{k,n} \gm5 + \gm5 D_{k,n} ] \psi_n
        \RR|_{A=0}\RR) J_\mu (m) \RR>                        \nn
&=& - \frac{i}{2} \sum_{m,n,k,i,j} n_{\sigma} m_{\lambda}
    \Bigl< \bpsi_k [ K_{\nu}(k,k-n) \gm5 + \gm5 K_{\nu}(k,k-n) ] \psi_{k-n}
\nn
&& \hspace{25mm}
         \bpsi_{m+i} K_{\mu}(i,i-j) \psi_{m+i-j} \Bigr>        \nn
&=& \frac{i}{2} \sum_{m,n,k,i,j} n_{\sigma} m_{\lambda} \Bigl\{ \nn
& &  \tr \LL( D^{-1}_{m+i-j-k} [ K_{\nu}(k,k-n) \gm5 + \gm5 K_{\nu}(k,k-n) ]
              D^{-1}_{k-n-m-i} K_{\mu}(i,i-j) \RR)      \nn
& & - \tr \LL( D^{-1}_{-n} [ K_{\nu}(k,k-n) \gm5 + \gm5 K_{\nu}(k,k-n) ] \RR)
      \tr \LL( D^{-1}_{-j} K_\mu(i,i-j) \RR)  \Bigr\}  \nn
&=& 0                                        \nonumber
\eea
where Eqs. (\ref{eq:gwr}) and (\ref{eq:JV}) have been used in the first
and the second equalities. From Eq. (\ref{eq:Kmu}), we see that
$ [ K_{\nu}(k,k-n) \gm5 + \gm5 K_{\nu}(k,k-n) ] $ is proportional to
$ \gm5 $, then using the trace identity Eq. (\ref{eq:gm5}) to give zero
in the above expression. Similarly, we show that $ P_4 $ also enters
(\ref{eq:I2'}) to give zero.

Finally only $ P_1 $ remains in Eq. (\ref{eq:I2'}), i.e.,
\bea
I_2 &=&  - 2 \sum_{m,n} n_{\sigma} m_{\lambda} P_1 \nn
    &=&  2 \sum_{m,n} n_{\sigma} m_{\lambda}
            \LL< (\bpsi DR\gm5 D)_n \psi_n J_{\mu}(m) J_{\nu}(0) \RR> \nn
    &=&  \sum_{m,n,k,i,j,s,t} n_{\sigma} m_{\lambda}
       \bigl< \bpsi_{k} ( D_{k-n}\gm5 + \gm5 D_{k-n} ) \psi_n \cdot \nn
    & &  \hspace{20mm} \bpsi_{m+i} K_{\mu}(i,i-j) \psi_{m+i-j}
          \bpsi_{s} K_{\nu}(s,s-t) \psi_{s-t}             \bigr>
\label{eq:I21}
\eea
where Eqs. (\ref{eq:gwr}) and (\ref{eq:JV}) have been used.
After the fermion fields are contracted, Eq. (\ref{eq:I21}) becomes
\BA
I_2 = I_a + I_b + I_c + I_d
\EA
where
\bea
I_a &=&  -\sum_{m,n,i,j,s} n_{\sigma} m_{\lambda}
    \tr[ \gm5 D^{-1}_{n-m-i} K_{\mu}(i,i-j) D^{-1}_{m+i-j-s} K_{\nu}(s,n) ]
\\
I_b &=&  -\sum_{m,n,i,s,t} n_{\sigma} m_{\lambda}
    \tr[ \gm5 D^{-1}_{n-s} K_{\nu}(s,s-t) D^{-1}_{s-t-m-i} K_{\mu}(i,n-m) ]
\\
I_c &=&  -\sum_{m,n,k,i,j,s,t} n_{\sigma} m_{\lambda}
       \tr[ \gm5 D_{k-n} D^{-1}_{n-m-i} K_{\mu}(i,i-j)
\nn
&& \hspace{40mm}
          D^{-1}_{m+i-j-s} K_{\nu}(s,s-t) D^{-1}_{s-t-k} ]
\\
I_d &=&  -\sum_{m,n,k,i,j,s,t} n_{\sigma} m_{\lambda}
         \tr[ \gm5 D_{k-n} D^{-1}_{n-s} K_{\nu}(s,s-t)
\nn
&& \hspace{40mm}
              D^{-1}_{s-t-m-i} K_{\mu}(i,i-j) D^{-1}_{m+i-j-k} ]
\eea
where those terms involving trace operation on products of one $ \gm5 $
and less than four $ \gamma_\mu's $ matrices are zero and have been dropped.
Using the Fourier transforms
\bea
D^{\pm 1}_n &=& \int_{-\pi}^{\pi} \frac{d^4 p}{(2\pi)^4} e^{ip \cdot n}
                                                         D^{\pm 1 } (p) \\
K_{\mu}(m,m-l) &=& \int_{-\pi}^{\pi} \frac{d^4 p}{(2\pi)^4}
                   \int_{-\pi}^{\pi} \frac{d^4 p'}{(2\pi)^4}
                   e^{ip \cdot l} e^{-ip'\cdot (m-l)} K_{\mu} (p,p+p')
\eea
and the identity
\bea
\int_{-\pi}^{\pi} \frac{d^4 p}{(2\pi)^4} \sum_n n_\sigma e^{ip \cdot n} f(p)
= \int_{-\pi}^{\pi} \frac{d^4 p}{(2\pi)^4} \delta^4(p) i
                    \frac{\partial}{\partial p_\sigma} f(p)
\eea
we obtain
\BA
I_a &=&  -\sum_{m,n,l,j,s} n_{\sigma} m_{\lambda} \int_{\{p_n\}}
   e^{ip_1(n-m-l)} e^{ip_2 l-ip_3 j} e^{ip_4(m+j-s)} e^{ip_5 s-ip_6 n}
\nn
&& \hspace{30mm}
   \tr[ \gm5 D^{-1}(p_1) K_{\mu}(p_2,p_3) D^{-1}(p_4) K_{\nu}(p_5,p_6)]
\nn
&=&  -\sum_{m,n,l,j,s} n_{\sigma} m_{\lambda} \int_{p_1 p_3 p_6}
        e^{ip_1(n-m)} e^{ip_3 m} e^{-ip_6 n}
\nn
&& \hspace{30mm}
     \tr[\gm5 D^{-1}(p_1)K_{\mu}(p_1,p_3)D^{-1}(p_3)K_{\nu}(p_3,p_6)]
\nn
&=&
    \int_{p_a p_b} \delta(p_a) \delta(p_b)
        \partial^{(a)}_{\lambda} \partial^{(b)}_{\sigma}
        \int_p \tr[ \gm5 D^{-1}(p+p_b) K_{\mu}(p+p_b,p+p_a+p_b)
\nn
&& \hspace{30mm}
                    D^{-1}(p+p_a+p_b) K_{\nu}(p+p_a+p_b,p)]
\label{eq:Iaa}
\EA
The differentiation with respect to $p_a$ and $p_b$ will generate many terms
in the last expression. However, due to the identity
(\ref{eq:gm5}), only those terms containing the product
$ \gm5 \gamma_1 \gamma_2 \gamma_3 \gamma_4 $ and its permutations
can have nonzero contributions, thus the final result is proportional to
$ \tr( \gm5 \gamma_\mu \gamma_\nu \gamma_\lambda \gamma_\sigma ) =
4 \epsilon_{\mu\nu\lambda\sigma} $.
Then it is obvious to see that those terms containing
$\partial^{(a)}_{\lambda}\partial^{(b)}_{\sigma}D^{-1}(p+p_a+p_b)$,
or $\partial^{(a)}_{\lambda}\partial^{(b)}_{\sigma}K_{\nu}(p+p_a+p_b,p)$
must vanish since they are symmetric in $ \lambda $ and $ \sigma $.
Furthermore, the terms containing $ \partial^{(a)}_{\lambda} K_{\mu}$,
$ \partial^{(b)}_{\sigma} K_{\mu}$, $ \partial^{(a)}_{\lambda} K_{\nu}$,
$ \partial^{(b)}_{\sigma} K_{\nu}$, or
$\partial^{(a)}_{\lambda}\partial^{(b)}_{\sigma}K_{\mu}(p+p_b,p+p_a+p_b)$
would become zero after integrations over
$ p_a $ and $ p_b $ with the delta functions in $ (\ref{eq:Iaa}) $, due to
the following identities ( proved in Appendix B )
\BA
\label{eq:K1}
\partial'_{\lambda} K_{\mu}(p,p+p')|_{p'=0} &=&
    \LL({i\over 2}\partial_{\mu}\partial_{\lambda}-
        {1\over 2}\delta_{\mu\lambda} \partial_{\mu}\RR)D(p)
\\
\label{eq:K2}
\partial'_{\lambda} K_{\mu}(p+p',p)|_{p'=0} &=&
    \LL({i\over 2}\partial_{\mu}\partial_{\lambda}+
        {1\over 2}\delta_{\mu\lambda} \partial_{\mu}\RR)D(p)
\EA
which are symmetric in $ \mu $ and $ \lambda $.
Hence, the only nonzero term in (\ref{eq:Iaa}) is
\bea
I_a &=& \int_p \tr \Bigl( \gm5 [ \partial_\sigma D^{-1}(p) ] K_{\mu}(p,p)
                   [ \partial_\lambda D^{-1}(p) ] K_{\nu}(p,p) \Bigr)  \nn
    &=& - \int_p \tr \Bigl( \gm5 [ \partial_\sigma D^{-1}(p) ]
                                 [ \partial_\mu D(p) ]
                                 [ \partial_\lambda D^{-1}(p) ]
                                 [ \partial_\nu D(p) ]          \Bigr)
\label{eq:Iaaa}
\eea
where the identity ( proved in Appendix B )
\BE
\label{eq:K3}
K_{\mu}(p,p) = i \partial_\mu D(p)
\EE
has been used.
By performing the same analysis on $ I_b $, $ I_c $
and $ I_d $, we obtain
\BA
I_b &=& - \sum_{m,n,l,s,t} n_{\sigma} m_{\lambda} \int_{\{p_n\}}
        e^{ip_1(n-s)} e^{ip_2 s-ip_3 t} e^{ip_4(t-m-l)} e^{ip_5 l-ip_6(n-m)}
\nn
&& \hspace{27mm}
        \tr[\gm5 D^{-1}(p_1)K_{\nu}(p_2,p_3)D^{-1}(p_4)K_{\mu}(p_5,p_6)]
\nn
&=&
    -\sum_{m,n}  n_{\sigma} m_{\lambda}  \int_{p_1 p_3 p_6}
        e^{ip_1 n} e^{-ip_3 m} e^{-ip_6(n-m)}
\nn
&& \hspace{27mm}
     \tr[\gm5 D^{-1}(p_1) K_{\nu}(p_1,p_3) D^{-1}(p_3) K_{\mu}(p_3,p_6)]
\nn
&=&
    \int_{p_a p_b} \delta(p_a) \delta(p_b)
        \partial^{(a)}_{\lambda} \partial^{(b)}_{\sigma}
        \int_p \tr[\gm5 D^{-1}(p+p_a+p_b) K_{\nu}(p+p_a+p_b,p)
\nn
&& \hspace{27mm}
                  D^{-1}(p)K_{\mu}(p,p+p_a)]
\nn
&=&
    0
\label{eq:Ibb}
\EA
where the zero is essentially due to the presence of $ D^{-1}(p) $
which does not depend on $ p_a $ or $ p_b $.
\BA
I_c &=& - \sum_{m,n,k,l,j,s,t} n_{\sigma} m_{\lambda} \int_{\{p_n\}}
        e^{ip_1(k-n)} e^{ip_2(n-m-l)} e^{ip_3 l-ip_4 j} e^{ip_5(m+j-s)}
        e^{ip_6 s-ip_7 t}
\nn
&&
        e^{ip_8(t-k)}
        \tr[ \gm5 D(p_1) D^{-1}(p_2) K_{\mu}(p_3,p_4) D^{-1}(p_5)
             K_{\nu}(p_6,p_7) D^{-1}(p_8) ]
\nn
&=&
    -\sum_{m,n}  n_{\sigma} m_{\lambda} \int_{p_1,p_2,p_4}
	e^{-ip_1 n}e^{ip_2(n-m)}e^{ip_4 m}
\nn
&& \hspace{22mm}
        \tr[\gm5 D(p_1)D^{-1}(p_2)K_{\mu}(p_2,p_4)D^{-1}(p_4)
	K_{\nu}(p_4,p_1)D^{-1}(p_1)]
\nn
&=&
    \int_{p_a p_b}\delta(p_a)\delta(p_b)
        \partial^{(a)}_{\lambda} \partial^{(b)}_{\sigma}
        \int_p \tr[\gm5 D(p) D^{-1}(p+p_b)
\nn
&&
	K_{\mu}(p+p_b,p+p_a+p_b)
	D^{-1}(p+p_a+p_b)K_{\nu}(p+p_a+p_b,p)D^{-1}(p)]
\nn
    &=& - \int_p \tr \Bigl( \gm5 [ \partial_\sigma D(p) ]
                                 [ \partial_\mu D^{-1}(p) ]
                                 [ \partial_\lambda D(p) ]
                                 [ \partial_\nu D^{-1}(p) ]          \Bigr)
\label{eq:Icc}
\EA
where the identity
\bea
D \partial_\mu D^{-1} + ( \partial_\mu D ) D^{-1} = 0
\eea
and Eqs. (\ref{eq:K1}), (\ref{eq:K2}) and (\ref{eq:K3}) have been used.

\BA
I_d &=& -\sum_{m,n,k,l,j,s,t} n_{\sigma} m_{\lambda} \int_{\{p_n\}}
        e^{ip_1(k-n)} e^{ip_2(n-s)} e^{ip_3 s-ip_4 t} e^{ip_5(t-m-l)}
        e^{ip_6 l-ip_7 j }
\nn
&&
        e^{ip_8(m+j-k)}
        \tr[\gm5 D(p_1) D^{-1}(p_2) K_{\nu}(p_3,p_4) D^{-1}(p_5)
             K_{\mu}(p_6,p_7) D^{-1}(p_8)]
\nn
&=&
     -\sum_{m,n}m_{\lambda}n_{\sigma} \int_{p_1 p_2 p_4}
	e^{-ip_1 n}e^{ip_2 n}e^{-ip_4 m}e^{ip_1 m}
\nn
&& \hspace{21mm}
        \tr[\gm5 D(p_1)D^{-1}(p_2)K_{\nu}(p_2,p_4)D^{-1}(p_4)
            K_{\mu}(p_4,p_1)D^{-1}(p_1)]
\nn
&=&
    \int_{p_a p_b}\delta(p_a)\delta(p_b)
        \partial^{(a)}_{\lambda} \partial^{(b)}_{\sigma}
        \int_p \tr[ \gm5 D(p+p_a) D^{-1}(p+p_a+p_b)
\nn
&& \hspace{21mm}
        K_{\nu}(p+p_a+p_b,p) D^{-1}(p) K_{\mu}(p,p+p_a) D^{-1}(p+p_a) ]
\nn
&=&
    \int_p \tr \Bigl(  \gm5 [\dl][\dsi][\dn][\dmi]    \nn
&& \hspace{20mm}      -\gm5 [\ds][\dni][\dm][\dli] \Bigr)
\nn
&=&
    -2 \int_p \tr \Bigl( \gm5 [\dmi][\dn][\dli][\ds] \Bigr)
\label{eq:Idd}
\EA
Therefore, summing $ I_a $, $ I_b $, $ I_c $ and $ I_d $ in
(\ref{eq:Iaaa}), (\ref{eq:Ibb}), (\ref{eq:Icc}) and (\ref{eq:Idd}),
and using Eqs. (\ref{eq:D0}), (\ref{eq:D0i}) and (\ref{eq:gm5}),
we obtain
\BA
\label{eq:I2a}
\hspace{-6mm}
I_2 &=& -4 \int_p \tr \Bigl( \gm5 [\dmi][\dn][\dli][\ds]  \Bigr) \nn
    &=& -4 \int_p \tr \Biggl[\gm5
        \partial_{\mu} \LL( {\ss{C}\over C^2} \RR)
        \partial_{\nu} \LL( {\ss{C} \over 1+r^2 C^2} \RR)
        \partial_{\lambda} \LL( {\ss{C}\over C^2} \RR)
        \partial_{\sigma} \LL( {\ss{C} \over 1+r^2 C^2} \RR) \Biggr]
\\
&=&   -4 \int_p \partial_{\mu} \ \tr \Biggl[ \gm5
        \LL({\ss{C}\over C^2} \RR)
        \partial_{\nu} \LL( { \ss{C} \over 1+r^2 C^2} \RR)
        \partial_{\lambda} \LL( { \ss{C} \over C^2} \RR)
        \partial_{\sigma}\LL({ \ss{C} \over 1+r^2 C^2} \RR)
                                                     \Biggr] \hspace{2mm}
\label{eq:I2b}
\EA
where the $ \partial_\mu $ operation in (\ref{eq:I2b})
produces (\ref{eq:I2a}), plus three terms which
are symmetric in $ \mu \nu $, $ \mu \lambda $, and $ \mu \sigma $,
respectively, hence neither of these three terms contributes to $ I_2 $.
Now we perform the momentum integral in (\ref{eq:I2b}) by, first removing
an infinitesimal ball $ B_\epsilon $ with center at the origin $ p = 0 $
and radius $ \epsilon $ from the Brillouin zone, then evaluating the
integral, and finally taking the limit $ \epsilon $ to zero, i.e.,
\BA
\label{eq:I2c}
I_2 &=& \frac{-4}{(2\pi)^4} \lim_{\epsilon \to 0}
        \int_{ \epsilon \le |p_\mu | \le \pi} {d^4 p} \
        \partial_{\mu} \Biggl \{
        \tr\Biggl[ \gm5
        \LL( {\ss{C} \over C^2} \RR)
        \partial_{\nu} \LL( {\ss{C} \over 1+r^2 C^2} \RR) \cdot
\nn
&& \hspace{50mm} \partial_{\lambda}\LL({\ss{C}\over C^2}\RR)
                 \partial_{\sigma}\LL({\ss{C} \over 1+r^2 C^2}\RR)
                                                    \Biggr] \Biggr\}
\EA
Then according to the Gauss theorem, the volume integral over the
Brillouin zone ( a four dimensional torus due to the periodic boundary
conditions ) with the ball $ B_\epsilon $ removed can be expressed as
a surface integral on the surface $ S_\epsilon $ of the ball $ B_\epsilon $,
provided that $ C_\mu (p) $ is nonzero for $ \epsilon \le |p_\mu | \le \pi $
( i.e., free of species doubling ) such that the integrand in
(\ref{eq:I2c}) is well defined. So, (\ref{eq:I2c}) becomes
\BA
I_2 &=& - \frac{1}{4 \pi^4} \lim_{\epsilon \to 0}
         \int_{S_\epsilon} {d^3 s} \ n_{\mu}
         \tr \Biggl[
         \gm5
         \LL( \frac{ \ss{C} }{C^2} \RR)
         \partial_{\nu} \LL( \frac{ \ss{C} }{1+r^2 C^2}  \RR) \cdot
\nn
&& \hspace{45mm} \partial_{\lambda} \LL( \frac{ \ss{C} } { C^2}    \RR)
                 \partial_{\sigma} \LL( \frac{\ss{C} }{ 1+r^2 C^2} \RR)
                                                                  \Biggr]
\EA
where $ n_\mu $ is the $\mu$-th component of the outward normal vector on
the surface $ S_\epsilon $.
Since we have assumed that $ C_\mu(p) \to p_\mu $ as $ p \to 0 $,
we can set $ C_\mu (p) = p_\mu $ on the surface $ S_\epsilon $
and obtain
\BA
I_2 &=&  - \frac{1}{4 \pi^4}
         \lim_{\epsilon \to 0} \int_{S_\epsilon}  {d^3 s} \ n_{\mu}
         \tr \Biggl[\gm5
         \LL(\ss{p}\over p^2\RR)\LL({\gamma_{\nu}\over 1+r^2 p^2}\RR)
         \LL(\gamma_{\lambda}\over p^2\RR)
         \LL({\gamma_{\sigma}\over 1+r^2 p^2}\RR)\Biggr]
\nn
&=&    - \frac{1}{\pi^4} \epsilon_{\mu\nu\lambda\sigma}
         \lim_{\epsilon \to 0} \int_{S_\epsilon} {d^3 s} \ n_\mu
         \frac{p_{\mu}}{p^4(1+r^2 p^2)^2}
\EA
where we have used the property
\BE
  \int_{S_\epsilon} {d^3 s} \ n_{\mu} p_{\nu} f(p^2)
= \delta_{\mu\nu} \int_{S_\epsilon} {d^3 s} \ n_{\mu} p_{\mu} f(p^2)
\EE
Finally, we have
\BA
I_2 &=& 
       - \frac{ \epsilon_{\mu\nu\lambda\sigma}}{\pi^4}
        \lim_{\epsilon \to 0} \frac{1}{(1+r^2 \epsilon^2)^2}
        \int^{2\pi}_{0} d\phi
        \int^{\pi}_0 d\theta_2 \sin \theta_2
        \int^{\pi}_0 d\theta_1 ( - \sin^2\theta_1 \cos^2\theta_1)
\nn
&=&
    \frac{1}{2\pi^2} \epsilon_{\mu\nu\lambda\sigma}
\label{eq:const}
\EA
This completes the task of evaluating
$I_2={\mathcal H}[ \ 2 \ \tr(\gm5 RD) \ ]$ for $ R = r \Id $,
where (\ref{eq:const}) is one of the main results of this paper.
Although (\ref{eq:const}) has been derived for the infinite
lattice ( a very large lattice with periodic boundary conditions ),
it is reasonable to expect that it also holds for finite lattices
with periodic boundary conditions and with even number of sites in
each direction, since the integral in (\ref{eq:I2c}) is essentially a
topological invariant quantity.
Implications of Eq. (\ref{eq:const}) have been discussed in section 2.


\section{Conclusions and Discussions}
\noindent

In this paper, we have evaluated the axial anomaly for the Ginsparg-Wilson
fermion operator $ D = D_c ( \Id + r D_c )^{-1} $. For any chirally
symmetric $ D_c $ which in the free fermion limit is free of species
doubling and behaves like $ i \gamma_\mu p_\mu $ as $ p \to 0 $,
the axial anomaly for $ U(1) $ lattice gauge theory with single fermion
flavor is
\BA
\hspace{-6mm} r \ \tr[ \gm5 D(x,x) ] &=&
   \frac{e^2}{32 \pi^2} \epsilon_{\mu\nu\lambda\sigma}
       F_{\mu\nu}(x) F_{\lambda\sigma}(x+\hat\mu+\hat\nu) \nn
&+& \mbox{ higher order  and/or  non-perturbative terms }
\hspace{2mm}
\label{eq:anomaly_r}
\EA
where the field tensor $ F_{\mu\nu} (x) $ on the lattice is defined in
Eq. (\ref{eq:Fuv}). The $ F \tilde{F} $ term is $r$-invariant and
has the correct continuum limit. As shown in section 2, the higher order
and/or non-perturbative terms might have significant impacts to the index
of $D$. For smooth background gauge configurations
[ e.g., Eq. (\ref{eq:U1})-(\ref{eq:U4}) ] with integer topological charge
(\ref{eq:ntop_lat}), if the axial anomaly satisfying Eq. (\ref{eq:ax_lat}),
then the sum of these higher order and/or non-perturbative terms over all
sites of the lattice is shown to be an integer multiple of the
$ F \tilde{F} $ term and this leads to the emergence of an integer
functional, $ c[D] $,
which is called topological characteristics of $ D $ in
ref. \cite{twc98:9a,twc98:10a}. In general, $ c[D] $ incorporates
all kinds of contributions from all fermionic modes.
Due to the $(S,T)$-invariance of the zero modes \cite{twc98:9a},
we have [ Eq. (\ref{eq:ASL}) in section 2 ]
\BA
N_{-} - N_{+} = c[D] \ \frac{e^2}{32 \pi^2} \sum_x
                     \epsilon_{\mu\nu\lambda\sigma}
                F_{\mu\nu}(x) F_{\lambda\sigma}(x+\hat\mu+\hat\nu)
\label{eq:index_thm_ST}
\EA
where $ c[D]=c[D_c] $ is invariant for any $ S $ and $ T $ in the general
GW relation, Eq. (\ref{eq:gwr_gen}). Then for topologically proper
$ D ( D_c ) $, i.e., $ c[D]=1 $, the Atiyah-Singer index theorem
can be realized on a finite ( infinite ) lattice for smooth background
gauge fields. Now it is obvious that the $ F \tilde{F} $ term
in Eq. (\ref{eq:anomaly_r}) must be also $(S,T)$-invariant, otherwise it
would be contrary to Eq. (\ref{eq:index_thm_ST}). Hence, we have
\BA
 \frac{1}{2} \ \tr \{ \gm5 [(S+T)D]_{x,x} \} &=&
 \frac{e^2}{32 \pi^2} \epsilon_{\mu\nu\lambda\sigma}
         F_{\mu\nu}(x) F_{\lambda\sigma}(x+\hat\mu+\hat\nu) \nn
  &+& \mbox{ higher order  and/or  nonperturbative terms. } \nn
\label{eq:anomaly_ST}
\EA
Then one can deduce the following result in the continuum limit,
\BA
 \frac{1}{2} \ \tr \{ \gm5 [(S+T)D]_{x,x} \} &=&
   c[D] \ \frac{e^2}{32 \pi^2} \epsilon_{\mu\nu\lambda\sigma}
                     F_{\mu\nu}(x) F_{\lambda\sigma}(x).
\label{eq:anomaly_STC}
\EA
Now the limit ( $ S,T \to 0 $ ) in Eq. (\ref{eq:anomaly_STC}) can be
safely taken. Since the limit $ S,T \to 0 $ is the chiral limit where
$ D = D_c $ and the GW chiral symmetry breaking
[ the RHS of Eq. (\ref{eq:gwr}) ] is completely turned off,
we conclude that the GW relation indeed does not play the crucial role
to fix the axial anomaly of $ D $ in the continuum limit.
This is in agreement with the conclusion of ref. \cite{twc98:9a}.
The crucial point for $ D $ to have the correct axial anomaly and
to realize the Atiyah-Singer index theorem in the continuum limit
is the existence of a topologically proper
$ D_c $ which also satisfies the properties mentioned above, or
in general, the constraints (a)-(e) given in ref. \cite{twc98:9a}.
Then any GW fermion operator $ D $ constructed by the general solution
$ D = D_c ( \Id + R D_c )^{-1} $ will have the desired topological
properties. The role of the chiral symmetry breaking transformation
(\ref{eq:gen_sol}) is to bypass the Nielson-Ninomiya no-go theorem
such that $ D $ can be constructed to be local, free of species doubling
and well defined for any gauge configurations, while the essential chiral
physics of $ D_c $ is preserved under this transformation.
Therefore, for practical computations on a finite lattice ( with finite
lattice spacings ), one must keep $ ( S, T ) $ finite as well as using
a topologically proper $ D ( D_c ) $ such that the axial anomaly
could agree with the Chern-Pontryagin density in continuum, though
the index is equal to the topologcial charge for any $ ( S, T ) $.

\appendix

\section{ }

\noindent
In this appendix, we explicitly show that
\bea
\label{eq:identity_ch}
 {\mathcal H'} [ \epsilon_{\alpha\beta\gamma\delta} F_{\alpha\beta} (x)
                     F_{\gamma\delta}(x+\hat\alpha+\hat\beta) ]
={\mathcal H'} [ \epsilon_{\alpha\beta\gamma\delta} F_{\alpha\beta} (x)
                     F_{\gamma\delta}(x) ]
= 8 \epsilon_{\mu\nu\lambda\sigma}
\eea
where $ {\mathcal H'} $ is defined as
\bea
\label{eq:H1}
 {\mathcal H'} [ O ] = \sum_{x,y} x_{\sigma} y_{\lambda}
 {\delta\over\delta A_{\nu}(0)}{\delta\over\delta A_{\mu}(y)}[ O(x) ]
\eea
which is similar to $ {\mathcal H} $ defined in Eq. (\ref{eq:HO}) but
without imposing the gauge fields to zero after the differentiations
with respect to the gauge fields.

First, we derive the field tensor for the $ U(1) $ gauge theory on the
lattice. A plaquette on the $ \hat\mu-\hat\nu $ plane is defined as
\bea
\label{eq:W11}
P_{\mu\nu} (x) &=&
  U_{\mu}(x) U_{\nu}(x+\hat\mu)
  U^{\dagger}_{\mu}(x+\hat\nu) U^{\dagger}_{\nu}(x) \nn
&=& \exp \Bigl\{ i e a \left[  A_{\nu}(x+\hat\mu)-A_{\nu}(x)
                             -A_{\mu}(x+\hat\nu)+A_{\mu}(x) \right] \Bigr\}
\eea
and its expansion up to $ e^2 $ is
\BA
\label{eq:W11e}
&&  1 + i e a \Bigl[ A_{\nu}(x+\hat\mu)-A_{\nu}(x)-
                     A_{\mu}(x+\hat\nu)+A_{\mu}(x)   \Bigr]
\nn
&&
    -{1\over 2} e^2 a^2 \Bigl[
        A_{\nu}(x+\hat\mu)-A_{\nu}(x)-A_{\mu}(x+\hat\nu)+A_{\mu}(x)
    \Bigr]^2
\nn
&\simeq&
    1 + i e a^2 \Bigl[\partial_{\mu}A_{\nu} - \partial_{\nu}A_{\mu}\Bigr]
        -{1 \over 2} e^2 a^4
        \Bigl[ \partial_{\mu}A_{\nu} - \partial_{\nu}A_{\mu} \Bigr]^2
        + O( e^3, a^6)
\EA
Then the real part of the sum of all plaquettes, i.e.,
\BA
\frac{1}{e^2} \sum_{x} \sum_{\nu < \mu }
                         \mbox{Re} [ 1- P_{\mu\nu}(x)],
\EA
goes to
\BA
{1 \over 4} \sum_{x,\mu,\nu} a^4
\Bigl[ \partial_{\mu}A_{\nu} - \partial_{\nu}A_{\mu} \Bigr]^2
\EA
in the continuum limit, thus agrees with the action of QED.
Hence, we can identify the field tensor on the lattice to be
\BA
\label{eq:Fuv_a}
F_{\mu\nu}(x) &=&
\frac{1}{a} \left[ A_{\nu}(x+\hat\mu)-A_{\nu}(x)
                  -A_{\mu}(x+\hat\nu)+A_{\mu}(x) \right] \\
&=& \frac{1}{i e a^2 } \log [ P_{\mu\nu} (x) ]
\label{eq:Fuv_b}
\EA
We note that on a finite lattice with periodic boundary conditions,
and for background gauge fields with nonzero topological charge,
some of the link variables at the boundary need modifications
[ Eqs. (\ref{eq:U2}) and (\ref{eq:U4}) ] such that the field tensors
are continuous on the torus.

Using Eqs. (\ref{eq:H1}) and (\ref{eq:Fuv_a}), we obtain
\BA
\label{eq:Fuv1}
& & {\mathcal H'} [  \epsilon_{\alpha\beta\gamma\delta} F_{\alpha\beta} (x)
                     F_{\gamma\delta}(x) ] \nn
&=& \sum_{x,y} \sum_{\alpha\beta\gamma\delta}  x_{\sigma} y_{\lambda}
    {\delta\over\delta A_{\nu}(0)}{\delta\over\delta A_{\mu}(y)} \Bigl\{
        \epsilon_{\alpha\beta\gamma\delta}
\nn
&&
      \Bigl[ A_{\beta}(x-\hat\alpha)A_{\delta}(x-\hat\gamma)
            -A_{\alpha}(x-\hat\beta)A_{\delta}(x-\hat\gamma)
\nn
&&
             -A_{\beta}(x-\hat\alpha)A_{\gamma}(x-\hat\delta)
             +A_{\alpha}(x-\hat\beta)A_{\gamma}(x-\hat\delta)
      \Bigr]
      \Bigr\}
\nn
&=&
    \sum_{x,y} x_{\sigma} y_{\lambda}
    {\delta\over\delta A_{\nu}(0)}{\delta\over\delta A_{\mu}(y)} \Bigl\{
\nn
&&
        (\epsilon_{\sigma\nu\lambda\mu}-\epsilon_{\nu\sigma\lambda\mu}
        -\epsilon_{\sigma\nu\mu\lambda}+\epsilon_{\nu\sigma\mu\lambda})
        A_{\nu}(x-\hat\sigma)A_{\mu}(x-\hat\lambda)
\nn
&&
       +(\epsilon_{\lambda\mu\sigma\nu}-\epsilon_{\mu\lambda\sigma\nu}
        -\epsilon_{\lambda\mu\nu\sigma}+\epsilon_{\mu\lambda\nu\sigma})
        A_{\mu}(x-\hat\lambda)A_{\nu}(x-\hat\sigma)
    \Bigr\}
\nn
&=&
\label{eq:ano}
    8 \epsilon_{\mu\nu\lambda\sigma}
\EA
where in the first and the second equalities, only those terms
which have nonzero contributions are retained.

For L\"uscher's topological charge density having the second
field tensor located at the site $ x+\hat\mu+\hat\nu $
rather than at $x$, we obtain
\BA
\label{eq:Fuv_2}
&& {\mathcal H'} [  \epsilon_{\alpha\beta\gamma\delta} F_{\alpha\beta} (x)
                     F_{\gamma\delta}(x+\hat\alpha+\hat\beta) ] \nn
&=& \sum_{x,y} \sum_{\alpha\beta\gamma\delta} x_{\sigma} y_{\lambda}
   {\delta\over\delta A_{\nu}(0)}{\delta\over\delta A_{\mu}(y)}\Bigl\{
        \epsilon_{\alpha\beta\gamma\delta}
\nn
&&
  \Bigl[ \
      A_{\beta}(x+\hat\alpha)A_{\delta}(x+\hat\alpha+\hat\beta+\hat\gamma)
     -A_{\beta}(x)A_{\delta}(x+\hat\alpha+\hat\beta+\hat\gamma)
\nn
&&
     -A_{\alpha}(x+\hat\beta)A_{\delta}(x+\hat\alpha+\hat\beta+\hat\gamma)
     +A_{\alpha}(x)A_{\delta}(x+\hat\alpha+\hat\beta+\hat\gamma)
\nn
&&
     -A_{\beta}(x+\hat\alpha)A_{\gamma}(x+\hat\alpha+\hat\beta+\hat\delta)
     +A_{\beta}(x)A_{\gamma}(x+\hat\alpha+\hat\beta+\hat\delta)
\nn
&&
     +A_{\alpha}(x+\hat\beta)A_{\gamma}(x+\hat\alpha+\hat\beta+\hat\delta)
     -A_{\alpha}(x)A_{\gamma}(x+\hat\alpha+\hat\beta+\hat\delta)
  \ \Bigr]
\nn
&=&
    \sum_{x,y} x_{\sigma} y_{\lambda}
    {\delta\over\delta A_{\nu}(0)}{\delta\over\delta A_{\mu}(y)}\Bigl\{
\nn
&&
  \epsilon_{\sigma\nu\lambda\mu}
      A_{\nu}(x+\hat\sigma)A_{\mu}(x+\hat\sigma+\hat\nu+\hat\lambda)
     +\epsilon_{\sigma\mu\lambda\nu}
      A_{\mu}(x+\hat\sigma)A_{\nu}(x+\hat\sigma+\hat\mu+\hat\lambda)
\nn
&-& \epsilon_{\sigma\mu\lambda\nu}
       A_{\mu}(x)A_{\nu}(x+\hat\sigma+\hat\mu+\hat\lambda)
     -\epsilon_{\lambda\mu\sigma\nu}
       A_{\mu}(x)A_{\nu}(x+\hat\lambda+\hat\mu+\hat\sigma)
\nn
&-& 
        \epsilon_{\nu\sigma\lambda\mu}
        A_{\nu}(x+\hat\sigma)A_{\mu}(x+\hat\nu+\hat\sigma+\hat\lambda)
       -\epsilon_{\mu\sigma\lambda\nu}
        A_{\mu}(x+\hat\sigma)A_{\nu}(x+\hat\mu+\hat\nu+\hat\lambda)
\nn
&+&
         \epsilon_{\mu\sigma\lambda\nu}
         A_{\mu}(x)A_{\nu}(x+\hat\mu+\hat\sigma+\hat\lambda)
       + \epsilon_{\mu\lambda\sigma\nu}
         A_{\mu}(x)A_{\nu}(x+\hat\mu+\hat\lambda+\hat\sigma)
\nn
&-&
        \epsilon_{\sigma\nu\mu\lambda}
        A_{\nu}(x+\hat\sigma)A_{\mu}(x+\hat\sigma+\hat\nu+\hat\lambda)
        -\epsilon_{\sigma\mu\nu\lambda}
        A_{\mu}(x+\hat\sigma)A_{\nu}(x+\hat\sigma+\hat\mu+\hat\lambda)
\nn
&+&
            \epsilon_{\lambda\mu\nu\sigma}
            A_{\mu}(x)A_{\nu}(x+\hat\lambda+\hat\mu+\hat\sigma)
           +\epsilon_{\sigma\mu\nu\lambda}
            A_{\mu}(x)A_{\nu}(x+\hat\sigma+\hat\mu+\hat\lambda)
\nn
&+&
        \epsilon_{\nu\sigma\mu\lambda}
        A_{\nu}(x+\hat\sigma)A_{\mu}(x+\hat\nu+\hat\sigma+\hat\lambda)
       +\epsilon_{\mu\sigma\nu\lambda}
        A_{\mu}(x+\hat\sigma)A_{\nu}(x+\hat\mu+\hat\sigma+\hat\lambda)
\nn
&-&
        \epsilon_{\mu\lambda\nu\sigma}
        A_{\mu}(x)A_{\nu}(x+\hat\mu+\hat\lambda+\hat\sigma)
       -\epsilon_{\mu\sigma\nu\lambda}
        A_{\mu}(x)A_{\nu}(x+\hat\mu+\hat\sigma+\hat\lambda)
 \Bigr\}
\nn
&=&
    8 \epsilon_{\mu\nu\lambda\sigma}
\EA
Among the eight lines of expressions at the second equality,
the second line vanishes due the cancellation of its two terms,
and the same happens to the fourth, the sixth and the eighth lines.
Then the remaining four lines add up to yield the final result.
This completes the proof of the identity (\ref{eq:identity_ch}).

\input{psfig.sty}

\section{ }

\noindent
In this appendix we derive some useful properties of the kernel $K_{\mu}$
of the vector current
\BA
J_{\mu}(n) = \sum_{m,l} \bpsi_{n+m} K_{\mu}(n,m,l;U) \psi_{n+l}.
\label{eq:Jmu}
\EA
These properties [ Eqs. (\ref{eq:KP1}) - (\ref{eq:KP3}) ]
are given in the appendix of Ginsparg and Wilson's original paper \cite{gwr}.
Here we present our derivation in details and correct a minor misprint
in ref. \cite{gwr}. As usual, the divergence of the vector current is
extracted from the change of the action under an infinitesimal local
transformation at the site $ n $,
\BAN
\psi_n \to \psi_n + \theta_n \psi_n  \\
\bpsi_n \to \bpsi_n - \theta_n \bpsi_n
\EAN
with the prescription
\BAN
{\mathcal A } \to {\mathcal A} + \theta_{n} \partial_{\mu} J_{\mu} (n)
\EAN
then we obtain
\BA
\partial_\mu J_{\mu} (n) = \sum_{m} [ \bpsi_{m} D_{mn} \psi_{n} -
                                       \bpsi_{n}  D_{nm} \psi_{m} ]
\label{eq:divJ_a}
\EA
where $ \partial_\mu J_{\mu} (n) $ is defined by the backward difference
\BA
\partial_\mu J_{\mu}(n)=
\sum_{\mu} \ [ J_{\mu} (n) - J_{\mu} ( n - \hat{\mu} ) \ ].
\label{eq:divJ_def}
\EA
Since we will turn off the gauge field after performing the
differentiations in Eq. (\ref{eq:I2}), we only need to derive all
properties of $ K_{\mu} $ in the free field limit where the action is
translational invariant $ D_{mn} = D_{m-n} $.
Then Eq. (\ref{eq:divJ_a}) can be rewritten as
\BA
\partial_\mu J_{\mu} (n) = \sum_{l} ( \bpsi_{n+l} D_{l} \psi_{n} -
                                      \bpsi_{n} D_{l} \psi_{n-l} ).
\label{eq:divJv}
\EA
and Eq. (\ref{eq:Jmu}) as
\BA
J_{\mu}(n) = \sum_{m,l} \bpsi_{n+m} K_{\mu}(m,m-l) \psi_{n+m-l}.
\label{eq:Jv}
\EA
To construct $ K_{\mu}(m,m-l) $ such that Eq. (\ref{eq:divJv})
can be reproduced with Eqs. (\ref{eq:divJ_def}) and (\ref{eq:Jv}),
the authors of ref. \cite{gwr} set
\BA
K_{\mu}(m,m-l) = L_{\mu}(m,l) \ D_l = \sign(l_{\mu}) \ f_{\mu}(m,l) \ D_l.
\label{eq:Kmu_a}
\EA
where $ f_{\mu}(m,l) $ is equal to the fraction of the shortest length
paths from $0$ to $l$ which pass through the link from $m-\hat\mu$ to $m$.
It is straightforward to verify that this definition of $ K_\mu $ leads to
Eq. (\ref{eq:divJv}). Using Eq. (\ref{eq:Jv}) and Eq. (\ref{eq:divJ_def}),
we obtain
\bea
\label{eq:JK}
\partial_\mu J_{\mu}(n) &=&
\sum_{\mu,m,l} [ \bpsi_{n+m} K_{\mu}(m,m-l) \psi_{n+m-l} 
\nn
&& \hspace{7mm}
    -\bpsi_{n-\hat{\mu}+m} K_{\mu}(m,m-l) \psi_{n-\hat{\mu}+m-l} ] 
\eea
The number of shortest length paths from $n+m-l$ to $n+m$ is
\BAN
N_l = \frac{ ( | l_1 | + | l_2 | + | l_3 | + | l_4 | ) ! }
           {   | l_1 | ! \ | l_2 | ! \ | l_3 | ! \ | l_4 | !   }.
\EAN
For a given set of positive intgers $ ( s_1, s_2, s_3, s_4 ) $,
and for all $ l $ with $ | l_\nu | = s_\nu, \nu=1, \cdots, 4 $, then
for $ l_{\mu} \ge 0 $, each one of these paths passing
through the link from $ n+m-\hat\mu $ to $ n+m $ contributes
$ \frac{1}{N_l} \bpsi_{n+m} D_l \psi_{n+m-l} $ to $ J_{\mu}(n) $;
while for $ l_{\mu} \le 0 $, a path passing through the link from $ n+m $
to $ n+m-\hat\mu $ contributes $ -\frac{1}{N_l} \bpsi_{n+m} D_l \psi_{n+m-l} $
to $ J_{\mu}(n) $. Hence for each $ l $, the contribution of each link
to the RHS of Eq. (\ref{eq:JK}) is
\BAN
\frac{1}{N_l} ( \bpsi_{n+m} D_l \psi_{n+m-l} -
                \bpsi_{n-\hat{\mu}+m} D_l \psi_{n-\hat{\mu}+m-l} ).
\EAN
Adding their contributions along each shortest length path would cancel in
pairs except the boundary terms
\BAN
\frac{1}{N_l} ( \psi_{n+l} D_l \psi_n - \psi_n D_l \psi_{n-l} ).
\EAN
The sum over all shortest length paths then cancels the factor
$ \frac{1}{N_l} $. Thus Eq. (\ref{eq:divJv}) is reproduced.

Next we prove two identities which are essential for deriving
Eqs. (\ref{eq:KP1}) - (\ref{eq:KP3}),
\BA
\sum_m L_{\mu}(m,l) &=& l_{\mu}.
\label{eq:Id1}
\\
\sum_m m_{\nu} L_{\mu}(m,l) &=&
\frac{ l_\mu ( l_\nu + \delta_{\mu\nu} ) }{2}.
\label{eq:Id2}
\EA
The summation over sites $m$ is defined as
\BAN
\sum_m \equiv \sum_{m_\mu }
              \sum_{m_\nu=0}^{ l_\nu} \sum_{m_\sigma=0}^{ l_\sigma }
              \sum_{m_\lambda=0}^{ l_\lambda }
\EAN
where the upper and lower limits of $ m_\mu $ depend on the sign of
$ l_\mu $. For $ l_\mu > 0 $, the summation of $ m_\mu $ is from $ 1 $
to $ l_\mu $, while for $ l_\mu < 0 $, from $ 0 $ to $ l_\mu + 1 $.
\BAN
\sum_{m_\mu} \equiv \left\{  \begin{array}{ll}
      \sum_{ m_\mu = 1 }^{ l_\mu }      &  \mbox{ if \ $ l_\mu > 0 $ } \\
      \sum_{ m_\mu = 0 }^{ l_\mu + 1 }  &  \mbox{ if \ $ l_\mu < 0 $ } \\
                             \end{array}
    \right.    \\
\EAN

\begin{figure}[ht]
\centerline{
    \psfig{file=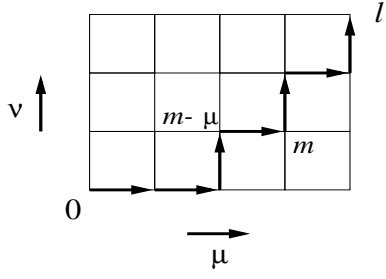,height=35mm,width=50mm,angle=-90} }
\caption{The portion of the lattice containing the shortest length paths
between $0$ and $l$. Only two directions ( $ \hat\mu $ and $ \hat\nu $ )
are shown on the plane while the other two directions
( $ \hat\sigma $ and $ \hat\lambda $ ) are orthogonal to the plane.
The solid line with arrows denotes the projection of one of the shortest
length paths onto the plane.}
\label{fig:paths}
\end{figure}

To prove the first identity, we observe that ( see Fig. 1 ) all shortest
length paths from $0$ to $l$ must go through one of the links pointing
in the $ \hat{\mu} $ direction with a fixed $ m_{\mu} $, i.e.,
they are all perpendicular to the hyperplane with fixed $ m_\mu $.
Therefore holding $ m_\mu $ fixed and summing over all other indices
( $ m_\nu, m_\sigma, m_\lambda $ ) of the fraction $ f_{\mu}(m,l) $
[ defined in Eq. (\ref{eq:Kmu_a}) ]
is equal to summing all probabilities for all shortest length paths
going through the hyperplane and hence it must equal to one.
Since there are $ l_{\mu} $ hyperplanes between
$0$ and $l_{\mu}$, after summing over $ m_{\mu} $, we obtain
Eq. (\ref{eq:Id1}),
\BAN
\sum_{m_\mu} \sum_{m_\nu=0}^{ l_\nu }
\sum_{m_\sigma=0}^{ l_\sigma } \sum_{m_\lambda=0}^{ l_\lambda }
\sign(l_\mu) f_{\mu}(m,l)
= \sum_{m_{\mu} } \sign(l_\mu) = l_\mu
\EAN
For the second identity, we first prove the case $ \mu = \nu $,
\BAN
& & \sum_{m_\mu} \sum_{m_\nu=0}^{ l_\nu }
    \sum_{m_\sigma=0}^{ l_\sigma } \sum_{m_\lambda=0}^{ l_\lambda }
    m_{\mu} L_{\mu}(m,l)
= \sum_{m_\mu} m_{\mu} \ \sign(l_\mu)      \\
&=& \left\{  \begin{array}{ll}
     ( 1 + 2 + \cdots + l_{\mu} )      &  \mbox{ if \ $ l_\mu > 0 $ } \\
     (-1 - 2 - \cdots - |l_\mu | + 1 ) &  \mbox{ if \ $ l_\mu < 0 $ } \\
             \end{array}
    \right.    \\
&=& \frac{  l_\mu (  l_\mu + 1 ) }{2}
\EAN
For $ \nu \ne \mu $, by symmetry, the fraction of shortest
length paths going through those links with fixed $ m_\mu $ and $ m_\nu $ is
( see Fig. 1 ),
\BAN
 \sum_{m_\sigma=0}^{ l_\sigma} \sum_{m_\lambda=0}^{ l_\lambda } f_\mu(m,l) =
      \frac{1}{ |l_\nu| + 1 }
\EAN
Then
\BAN
\sum_{m_{\mu}} \sum_{m_\nu=0}^{ l_\nu }
\sum_{m_\sigma=0}^{ l_\sigma } \sum_{m_\lambda=0}^{ l_\lambda }
     m_{\nu} L_{\mu}(m,l)
= \sum_{m_\mu} \sum_{m_\nu=0}^{ l_\nu }
    m_\nu \ \sign(l_\mu) \frac{1}{ | l_\nu | + 1 }
= \frac{ l_\mu \ l_\nu }{2}
\EAN
This completes the proof of the second identity.

With these two identities, we proceed to derive some useful properties of
$K_{\mu}$ in momentum space
\BAN
K_{\mu}(p,p+p') &=& \sum_{l,m} e^{-ipm} e^{i(p+p')(m-l)} K_{\mu}(m,m-l) \\
&=&
 \sum_{l,m} e^{-ipl} e^{ip'(m-l)} L_{\mu}(m,l) D_l.
\EAN
Since we only need the derivatives of $K_{\mu}$ with respect to
$p'_{\nu}$ evaluated at $p'=0$, we can expand the above equation
up to first order of $p'$
\BAN
K_{\mu}(p,p+p') &\approx& 
  \sum_{l,m} e^{-ipl} [1+ip'(m-l)] L_{\mu}(m,l) D_l \\
&=&
\sum_l e^{-ipl} l_{\mu} D_l \LL(1+i{p'_{\nu} l_{\nu} +p'_{\mu}\over 2}-ip'l\RR) \\
&=& i\partial_{\mu} \LL( 1+{p'_{\nu} \partial_{\nu}\over 2}
                          +{ip'_{\mu} \over 2} \RR) D(p)
\EAN
where $ D(p) = \sum_l e^{-ipl} D_l $ and Eqs. (\ref{eq:Id1}), (\ref{eq:Id2})
have been used. Therefore, we have
\BA
K_{\mu}(p,p) = i \partial_{\mu} D(p)
\label{eq:KP1}
\EA
and
\BA
\partial'_{\nu} K_{\mu}(p,p+p')|_{p'=0} =
    \LL( {i\over 2} \partial_{\mu} \partial_{\nu} -
         {1\over 2} \delta_{\mu\nu} \partial_{\mu} \RR) D(p)
\label{eq:KP2}
\EA
Similarly, we obtain
\BAN
K_{\mu}(p+p',p) &=& \sum_{l,m} e^{-i(p+p')m} e^{ip(m-l)} K_{\mu}(m,m-l) \\
&=& \sum_{l,m} e^{-ipl} e^{-ip'm} K_{\mu}(m,m-l) \\
&\approx&  \sum_{l,m} e^{-ipl} (1-ip'm) K_{\mu}(m,m-l) \\
&=& \sum_l e^{-ipl} l_{\mu} D_l \LL(1-i{p'_{\nu} l_{\nu} +p'_{\mu}\over 2}\RR)\\
&=& i\partial_{\mu} \LL( 1+{p'_{\nu}\partial_{\nu}\over 2}
                         -{ip'_{\mu}\over 2} \RR) D(p)
\EAN
and this leads to
\BA
\partial'_{\nu} K_{\mu}(p+p',p)|_{p'=0} =
    \LL( {i\over 2} \partial_{\mu} \partial_{\nu} +
         {1\over 2} \delta_{\mu\nu}\partial_{\mu} \RR) D(p)
\label{eq:KP3}
\EA
We note that the operator $ \partial_{\mu} $ in the second term of
Eq. (\ref{eq:KP2}) is missed in Eqs. (37) and (A6) of Ginsparg and
Wilson's original paper \cite{gwr}.
Equations (\ref{eq:KP1})-(\ref{eq:KP3}) are used in our
derivation of chiral anomaly in section 3.

\section{ }

In this appendix we prove the second equality of Eq. (\ref{eq:JVU})
[ or Eq. (\ref{eq:JV}) ] which has played an important
role in our derivation of axial anomaly in section 3.
The vector current in Eq. (\ref{eq:JVU}) is
\beq
\label{eq:J_def}
J_{\mu}(k,U) = \sum_{i,j} \bpsi_{k+i} K_{\mu}(k,i,j;U) \psi_{k+j}
= i\sum_{m,n}\bpsi_m{\delta\over\delta A_{\mu}(k)}D_{mn}(U)\psi_n
\eeq
An explicit realization of the kernel $K_{\mu}(k,i,j;U)$ is given in
ref. \cite{gwr} as
\beq
\label{eq:K_def}
K_{\mu}(k,i,j;U) = \sign((i-j)_{\mu}) f_{\mu}(k+i,k+j) D_{k+i,k+j}(U)
\eeq
where $f_{\mu}(k+i,k+j)$ is the fraction of the shortest length paths 
from $k+j$ to $k+i$ which pass through the link $ (k,k+\hat\mu)$.
Note that the vector current $ J_{\mu}(k,U) $ satisfying
Eq. (\ref{eq:divJ}) is not unique. But the kernel defined in
Eq. (\ref{eq:K_def}) leads to the second equality of
Eq. (\ref{eq:J_def}).
First we note that the ordered product of link variables
along one of the shortest length paths from $n$ to $m$, say, $P$, is
\beq
U_P(m,n) = \prod_{P} U(m,s) \cdots U(t,n)
\eeq
where $ U(m,s) $ denotes the link variable pointing from $ m $ to $ s $ with
the usual convention
$$
  U(m,m+\hat\mu) = U_{\mu}(m) = \exp[ i a e A_{\mu} (m) ].
$$
Then $ U_P(m,n) $ enters the action
$ {\mathcal A } = \sum_{m,n} \bpsi_m D_{mn}(U) \psi_n $
via the following gauge invariant product
\beq
\label{eq:AP}
  \bpsi_m  \Gamma(m,n)  U_P(m,n) \psi_n
\eeq
where $ \Gamma(m,n) $ is a matrix in the Dirac space and its
explicit form is irrelevant to our present discussion.
The normalized sum of (\ref{eq:AP}) over all shortest length paths
from $ n $ to $ m $ is equal to the term $ \bpsi_m D_{mn}(U) \psi_n $
in the action, i.e.,
\beq
\label{eq:DUP}
 D_{mn}(U) = \Gamma(m,n) \frac{1}{N_l} \sum_{P} U_P(m,n)
\eeq
where $ N_l $ is the total number of shortest length paths from
$ n $ to $ m $,
\beq
N_l = {(|l_1|+|l_2|+|l_3|+|l_4|)! \over |l_1|!\ |l_2|!\ |l_3|!\ |l_4|!},
\hspace{4mm} l = m - n
\eeq
Then using the following simple identity
\beq
 {\delta\over\delta A_{\nu}(k)} U_{\mu}(n) =
 i\delta_{\mu\nu}\delta_{kn}U_{\mu}(n)
\eeq
it is straightforward to obtain that, if $U_P(m,n)$ contains the link
$ U_{\mu}(k) $ ( or its hermitian conjugate ), then
the derivative of $D_{mn}(U)$ with respect to $A_{\mu}(k)$
yields $ i \ U_P(m,n) $ times a sign factor which depends on the
relative positions of $m$ and $n$ ( i.e. $-1$ for $ m_{\mu} > n_{\mu} $
but $+1$ for $ m_{\mu} < n_{\mu} $ [ see Eq. (\ref{eq:AP}) ] ); and
the derivative is zero if $ U_P(m,n) $ does not contain the link
$ U_{\mu}(k) $. That is
\beq
\hspace{-8mm}{\delta\over\delta A_{\mu}(k)} U_P(m,n) =
\left\{ \begin{array}{ll}
        -i \ \mbox{sign}( (m-n)_{\mu} ) U_P(m,n), &
         \mbox{ if $ U_{\mu}(k) \in  U_P(m,n) $} \\
         0, & \mbox{ otherwise } \\
         \end{array}
         \right.
\label{eq:dUPdA}
\eeq
Now multiplying both sides of Eq. (\ref{eq:dUPdA}) by $ \Gamma(m,n) $,
summing over all shortest length paths from $ n $ to $ m $, and dividing
by $ N_l $, we obtain
\bea
 i {\delta\over\delta A_{\mu}(k)} D_{mn}(U) &=&
 \mbox{sign}( (m-n)_{\mu} ) \Gamma(m,n) \frac{1}{N_l} \sum_{P'} U_{P'}(m,n) \nn
&=& \mbox{sign}( (m-n)_{\mu} ) f_{\mu}(m,n)
  \Gamma(m,n) \frac{1}{N_l^{'}} \sum_{P'} U_{P'}(m,n)
\label{eq:dDdA}
\eea
where Eq. (\ref{eq:DUP}) has been used on the LHS. The summation
$ P'$ on the RHS denotes the sum over the shortest length paths between
$ m $ and $ n $ which pass through the link $ ( k,k+\hat\mu) $, and
the total number of these paths is denoted by $ N_l^{'} $, and
$ f_{\mu}(m,n) = N_{l}^{'} / N_{l} $
is the fraction of the shortest length paths from $n$ to $m$
which pass through the link $ (k,k+\hat\mu) $.
Now we sandwich both sides of Eq. (\ref{eq:dDdA}) by $ \bpsi_m $ and
$ \psi_n $ and sum over $m$ and $n$. Then on the RHS of the
resulting equation, we can write $ m = k + i $ and $ n = k + j $
and replace summations over $ m $ and $ n $ by summations over
$ i $ and $ j $, since only those shortest length paths from $n$ to $m$
which pass through the link $ (k,k+\hat\mu) $ can have nonzero
contributuion, finally we have
\bea
& & i \sum_{m,n} \bpsi_{m} {\delta\over\delta A_{\mu}(k)} D_{mn}(U) \psi_n \nn
&=& \sum_{i,j} \bpsi_{k+i} \ \mbox{sign}( (i-j)_{\mu} ) \ f_{\mu}(k+i,k+j)
            \ D_{k+i,k+j}(U) \ \psi_{k+j}                    \\
&=& \sum_{i,j} \bpsi_{k+i} K_{\mu}(k,i,j;U) \psi_{k+j}
\eea
where
\beq
D_{k+i,k+j}(U)=\Gamma(k+i,k+j) \frac{1}{N_l^{'}} \sum_{P'} U_{P'}(k+i,k+j).
\eeq
This completes the proof of the second equality of Eq. (\ref{eq:JVU}).
%
%
%

\bigskip
\bigskip
\bigskip
\flushpar
{\bf Acknowledgement }
\bigskip

\noindent
This work was supported by the National Science Council, R.O.C.
under the grant number NSC88-2112-M002-016.

\vfill\eject

\end{document}